\documentclass[acmsmall]{acmart}

\usepackage{caption}
\usepackage{subcaption}
\usepackage{array}
\usepackage{threeparttable}
\usepackage{multirow}
\usepackage{enumitem}
\usepackage{booktabs}
\usepackage{soul}

\AtBeginDocument{%
  \providecommand\BibTeX{{%
    \normalfont B\kern-0.5em{\scshape i\kern-0.25em b}\kern-0.8em\TeX}}}

\copyrightyear{2022}
\acmYear{2022}
\setcopyright{acmcopyright}\acmConference[CHI '22]{CHI Conference on Human Factors in Computing Systems}{April 29-May 5, 2022}{New Orleans, LA, USA}
\acmBooktitle{CHI Conference on Human Factors in Computing Systems (CHI '22), April 29-May 5, 2022, New Orleans, LA, USA}
\acmPrice{15.00}
\acmDOI{10.1145/3491102.3502089}
\acmISBN{978-1-4503-9157-3/22/04}

\begin{document}

\title[To Trust or to Stockpile]{To Trust or to Stockpile: Modeling Human-Simulation Interaction in Supply Chain Shortages}

\author{Omid Mohaddesi}
\orcid{0000-0002-6245-4003}
\affiliation{%
  \institution{Northeastern University}
  \streetaddress{360 Huntington Ave.}
  \city{Boston}
  \state{Massachusetts}
  \country{USA}
  \postcode{02115}
}
\email{mohaddesi.s@northeastern.edu}

\author{Jacqueline Griffin}
\affiliation{%
  \institution{Northeastern University}
  \streetaddress{360 Huntington Ave.}
  \city{Boston}
  \state{Massachusetts}
  \country{USA}
  \postcode{02115}
}
\email{ja.griffin@northeastern.edu}

\author{Ozlem Ergun}
\affiliation{%
  \institution{Northeastern University}
  \streetaddress{360 Huntington Ave.}
  \city{Boston}
  \state{Massachusetts}
  \country{USA}
  \postcode{02115}
}
\email{o.ergun@northeastern.edu}

\author{David Kaeli}
\affiliation{%
  \institution{Northeastern University}
  \streetaddress{360 Huntington Ave.}
  \city{Boston}
  \state{Massachusetts}
  \country{USA}
  \postcode{02115}
}
\email{d.kaeli@northeastern.edu}

\author{Stacy Marsella}
\affiliation{%
  \institution{Northeastern University}
  \streetaddress{360 Huntington Ave.}
  \city{Boston}
  \state{Massachusetts}
  \country{USA}
  \postcode{02115}
}
\email{s.marsella@northeastern.edu}

\author{Casper Harteveld}
\affiliation{%
  \institution{Northeastern University}
  \streetaddress{360 Huntington Ave.}
  \city{Boston}
  \state{Massachusetts}
  \country{USA}
  \postcode{02115}
}
\email{c.harteveld@northeastern.edu}

\renewcommand{\shortauthors}{Mohaddesi, et al.}

\begin{abstract}
Understanding decision-making in dynamic and complex settings is a challenge yet essential for preventing, mitigating, and responding to adverse events (e.g., disasters, financial crises). Simulation games have shown promise to advance our understanding of decision-making in such settings. However, an open question remains on how we extract useful information from these games. We contribute an approach to model human-simulation interaction by leveraging existing methods to characterize: (1) system states of dynamic simulation environments (with Principal Component Analysis), (2) behavioral responses from human interaction with simulation (with Hidden Markov Models), and (3) behavioral responses across system states (with Sequence Analysis). We demonstrate this approach with our game simulating drug shortages in a supply chain context. Results from our experimental study with 135 participants show different player types (hoarders, reactors, followers), how behavior changes in different system states, and how sharing information impacts behavior. We discuss how our findings challenge existing literature.
\end{abstract}




\maketitle



\section{Introduction}
When the COVID-19 pandemic started, many experts were concerned about shortages in medical equipment~\cite{ranney2020critical} and drugs~\cite{roos_experts_2020, rees_ema_2020}. Perhaps the last thing everyone could predict was a toilet paper shortage, which was a direct result of panic-buying and hoarding behavior of people~\cite{david2021did}. Understanding such behaviors has been the core focus of many psychological~\cite{arafat2020psychological} and supply chain research~\cite{sterman2015m}. While a toilet paper shortage may not sound very crucial, the same type of behavior can significantly impact the supply of many critical products, including pharmaceutical drugs. Therefore, it is essential to investigate the role of human behavior and decision-making in critical systems and especially in the presence of adverse events (e.g., natural disasters, health crises, financial crises).

Simulation games have been increasingly used as a research environment to study human decisions in critical contexts. In particular, methods such as gaming simulation~\cite{meijer2009organisation} and participatory simulation~\cite{guyot2006agent, anand2016validation}, and game-based simulation environments such as gamettes~\cite{mohaddesi_introducing_2020} have shown promise for advancing our understanding of human behavior. Researchers create these environments to validate and improve the underlying simulation by observing or modeling human behavior. However, a key challenge in modeling human behavior in such environments is the system's dynamic nature. While the system's state affects humans' behavior, this state can change due to human interaction with the simulation. In addition, researchers often use these environments to generate hypotheses and test them~\cite{meijer2009organisation}; hence, they create manipulations that may have unexpected effects on human behavior~\cite{gundry2019validity}. When coupled with a dynamic system, teasing out the effect of such manipulations becomes more complicated, making behavioral modeling in such environments challenging. We aim to extend existing research in behavioral modeling within game-based simulation environments by characterizing the human-simulation interaction.

To this end, we propose an approach to characterize (1) the system's states, (2) the behavioral responses from human participants interacting with the simulation, and (3) the interaction of humans with the simulation through analyzing their behavioral responses across the system's states while taking manipulations into account. We use existing methods for our characterization in each step. In addition, we use a gamette environment as described in our previous work~\cite{mohaddesi_introducing_2020} and conduct an experimental study to put human participants into the loop of an agent-based simulation, replicating a drug delivery supply chain. Our simulation models a shortage scenario capturing interdependencies in a supply chain. By characterizing the system, we identify different supply chain states for our participants' roles. We show how to characterize the state of such a high dimensional system by leveraging: (1) Principal Component Analysis (PCA) to reduce the dimensionality of our supply chain simulation and (2) hierarchical cluster analysis to find representative system's states. 

A gamette environment creates an authentic decision context for human players~\cite{mohaddesi_introducing_2020}. We use telemetry data provided by the gamette environment for modeling player behavior. According to a recent review~\cite{hooshyar2018data}, Hidden Markov Models (HMMs) are employed for player behavioral modeling, and are especially promising for modeling players' differences in their sequential decisions~\cite{bunian2017modeling}. Prior studies outside the realm of player modeling also utilized HMMs for analyzing sequential data~\cite{mor2020systematic} and modeling cognitive processes~\cite{kunkel2020hierarchical}. We use HMMs to characterize player response modes in their decision-making within our supply chain decision task. We further show how these response modes can be used for player profiling and identifying different player types by leveraging sequence analysis. We identify distinct decision patterns by comparing players' deviations from system recommendations in their ordering decisions. Finally, we analyze the sequences of response modes across the system's states to characterize the interaction of human participants with our supply chain simulation. In particular, we show how the behavior changes in different system's states. We also show how our manipulation of the environment (i.e., information sharing in the context of supply chains) impacts behavior. 

The contributions of this work are as follows:
\begin{itemize}
    \item We re-institute human-simulation interaction as a research avenue within HCI research with an emphasis on characterizing the interaction between the human and simulation, hence, taking the dynamic state of the system as well as environment manipulation into account for behavioral modeling. We do this by showing how to (1) characterize the states of a simulation environment and (2) characterize the behavioral responses of humans interacting with the simulation.
    \item We identify three player types (hoarders, reactors, followers) with distinct decision-making patterns in interaction with a supply chain simulation game replicating drug shortages.
    \item We provide evidence for the effect of information sharing on each player type through characterizing human-simulation interaction.
\end{itemize}

\section{Related Work}
In our review of the related research, we first summarize the studies around using games for involving humans in the simulation. We then look into modeling human behavior and decision-making, especially in the context of supply chain decisions. Finally, we review the studies focusing on player modeling and the methods used for modeling player behavior in games.

\subsection{Game-Based Simulation Environments}
The involvement of humans with simulations and using games as a medium to facilitate their interaction is an old but ongoing concept. At a high level, it is referred to as participatory approaches, which involve the inclusion of humans during the design, implementation, execution, and analysis of a simulation~\cite{barreteau2017participatory}. For example, Guyot and Honiden~\cite{guyot2006agent} introduced a specific form of a participatory approach called \textit{agent-based participatory simulation} which involves merging role-playing games (RPGs) and multi-agent systems (MAS). Such a combination aims to have human participants control some of the agents in the simulation rather than involving participants in the design process, which allows analyzing participants' behavior in detail. In particular, Anand et al.~\cite{anand2016validation} used this approach for validating agents in an agent-based simulation through engaging human participants with the simulation. Other work in this area involves using agent-based participatory simulation to improve students' understanding of different components in complex systems~\cite{rates2016promoting}. 

In our previous work~\cite{mohaddesi_introducing_2020}, we proposed a similar approach by introducing a game-based methodology called \textit{gamettes} for involving human participants in the simulation. The difference between our approach and agent-based participatory simulation is that gamettes target a short and specific part of the simulation instead of the whole simulation. We tested our approach in a supply chain simulation replicating the Beer Distribution Game and showed that the gamette could capture supply chains' expected behavioral patterns. Similarly, Meijer~\cite{meijer2009organisation} introduced gaming simulation as a research method for allowing humans to enact a role in a simulated environment. Gaming simulation has been used to test hypotheses and validate the simulation~\cite{van2017assessing}, and also expanded to use in engineering systems research~\cite{grogan2017gaming}. 

We refer to all these approaches (i.e., agent-based participatory simulation, gamettes, gaming simulation), which aim to incorporate human behavior with a simulation to collect data on how humans interact with a simulated environment, as ``game-based simulation environments''. In this work, we use gamettes as our approach for involving humans with a simulation through an experimental study replicating supply chain shortages. We use gamettes again as it has shown promise in our prior work for conducting behavioral experiments in the context of supply chains~\cite{mohaddesi_introducing_2020}. Here, however, we leverage this game-based simulation environment to advance our understanding of human decision-making in a dynamic system. 

\subsection{Behavioral Modeling}
Humans make decisions by reasoning about their complex environments. Behavioral modeling aims to improve the decision processes by investigating the patterns of reasoning and making sense of those patterns~\cite{pomerol1997artificial}. Researchers in various fields have dedicated years to understanding human decision-making. In the context of supply chains, operations researchers have attempted to model human behavior using behavioral research methods~\cite{tokar2010behavioural, kunc2016behavioral} or system dynamics~\cite{sterman1989modeling, sterman2015m}. Prior research in this area involve studies that aimed at modeling: the mental models of decision-makers, including models of rationality~\cite{narayanan2015decision, haines2010individual} and judgment~\cite{edali2016results, tokar2016exploring}; decision strategies~\cite{delhoum2009influence}; individual traits and demographics~\cite{turner2020results}; emotions~\cite{nienhaus2006human}; or decision patterns~\cite{sterman2015m}. In this study, we consider modeling ordering decisions by characterizing behavioral responses from human participants. Our goal is to find response modes that explain the decision patterns of our participants. 

Supply chains are dynamic systems, and modeling human behavior in this context can be regarded as modeling dynamic decision-making. Prior research used computer simulations referred to as ``microworlds'' to study dynamic decision-making in a laboratory setting~\cite{gonzalez2005use} or leveraged serious games for analyzing dynamic decisions~\cite{nonaka2016analysis}. However, using game-based simulation environments in a dynamic system adds a level of challenge to modeling behavior. Previous research has provided evidence for how only a slightly different individual behavior can impact the system, resulting in very different trajectories in the state that players find themselves in~\cite{sun2016modeling}. Therefore, we use an approach for behavioral modeling where we characterize the dynamic state of the system. Through such characterization, we provide an opportunity to zoom in on the interaction of human behavior with the system. Prior studies also investigated ways in which decision-making can be supported in dynamic settings~\cite{gonzalez2005decision}. In our experiment, we provide such decision support to our participants by providing them with order suggestions according to a base stock policy~\cite{snyder2011fundamentals}. We attempt to explain decision patterns by considering the participants' level of deviation from these order suggestions.

\subsection{Player Modeling}
As we rely on games as our medium to engage human participants with a simulation, we can leverage the research around player modeling to characterize our participants' behavior. Player modeling is aimed at understanding how players experience their interaction with the game by focusing on their patterns of behavior through modeling cognitive or affective states~\cite{yannakakis2013player}. Previous studies in the realm of player modeling used various methods for modeling and imitating player actions, including clustering techniques~\cite{drachen2016guns, drachen2012guns}, imitation learning~\cite{partlan2019player, mohaddesi2020learning}, neural networks~\cite{burelli2015adapting}, and Bayesian models~\cite{synnaeve2011bayesian, normoyle2021bayesian}. Other researchers used Hidden Markov Models (HMM) for modeling player actions in the game~\cite{bunian2017modeling, pfau2018towards}. HMMs have shown great promise, especially for analyzing sequential data~\cite{mor2020systematic} and modeling cognitive processes~\cite{kunkel2020hierarchical}. Therefore, we use HMM to model player response modes in their ordering decisions, specifically their deviations from order suggestions provided by the system. In addition, previous work demonstrated the potential of using sequence analysis in gameplay analysis~\cite{kleinman2020and}. We leverage such methods to analyze response modes' sequences and profile players. The literature on player modeling also includes studies that focus on clustering game states and players' responses while mapping state-response clusters for generating player models~\cite{bindewald2016clustering}. Our work is distinguished from these prior studies and existing player modeling research in the sense that our focus is on characterizing the human-simulation interaction. We do not model behavioral response in isolation. Rather, we characterize the interaction of human behavior by analyzing response modes over system states and focusing on how the interaction with the simulation environment forms human behavior.

\begin{figure}[ht]
    \centering
    \includegraphics[width=.5\columnwidth]{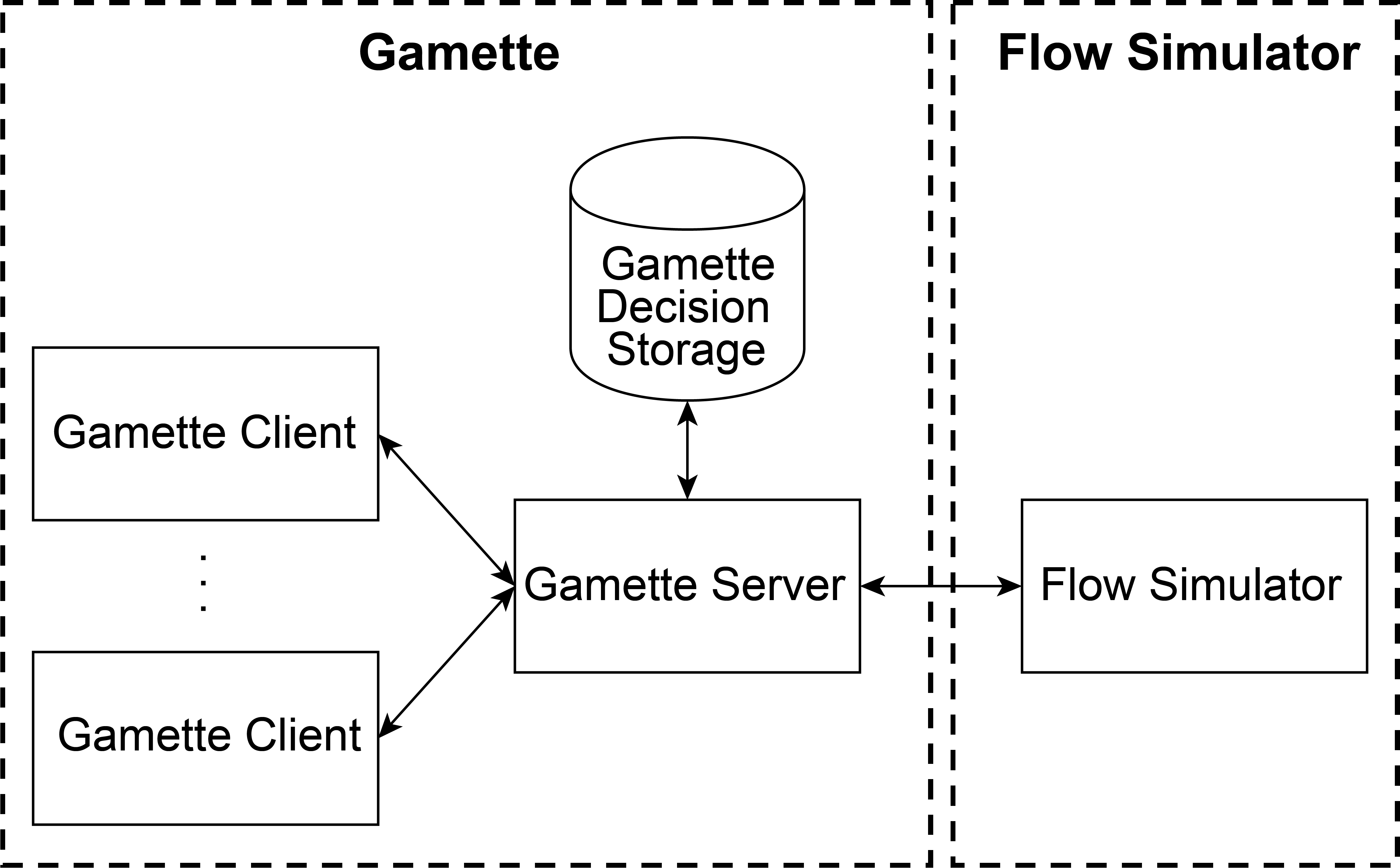}
    \caption{The integrated simulation framework.}\label{fig:simulationframework}
    \Description[Simulation architecture including gamette and Flow Simulator]{The architecture of the integrated simulation framework shows how the gamette environment is connected to the Flow simulator. Each gamette environment comprises a gamette server that communicates with Flow Simulator. Multiple gamette clients can communicate with the gamette server. Gamette server stores all game data in gamette decision storage.}
\end{figure}

\section{gamettes}
As described in our previous work~\cite{mohaddesi_introducing_2020}, gamettes is a methodology to capture behavioral aspects of human decision-making. Gamettes are short game-based scenarios where individual decision-makers are immersed in a specific situation and make decisions by responding to a dialog or taking action. The term gamette is a contraction of ``game'' with ``vignette''. Similar to a vignette, a gamette aims to provide a \textit{brief description} of a situation as well as to \textit{portray} someone. Here, we focus on modeling human behavior and decision-making in a drug delivery supply chain. For this, we use the integrated simulation framework proposed by~\cite{doroudi2018integrated}. This framework encompasses a \textit{Flow Simulator} for simulating the supply chain dynamics and a \textit{gamette} environment for engaging human decision-makers with the simulation through immersing them into a specific role and particular state of the supply chain. 

The Flow Simulator is the core of this framework that simulates the information and physical flow in the drug delivery supply chain over time. The decisions and actions taken by the agents of the system (i.e., manufacturers, wholesalers, and health centers) drive these information and physical flows. Such decisions and actions can be the result of running the Flow Simulator in a \textit{standalone mode}. In that case, the Flow Simulator simulates the evolution of the supply chain system with predefined policies informing decision-making---therefore, without any human agents. Conversely, the Flow Simulator can simulate by fetching information from the gamette clients, which capture decisions of human players. Here, the evolution of the supply chain system is a result of human input. Figure~\ref{fig:simulationframework} illustrates the architecture of our simulation framework and the interaction of its components.

\begin{figure}[ht]
    \centering
    \begin{subfigure}[t!]{.35\columnwidth}
        \centering
        \includegraphics[width=.9\columnwidth]{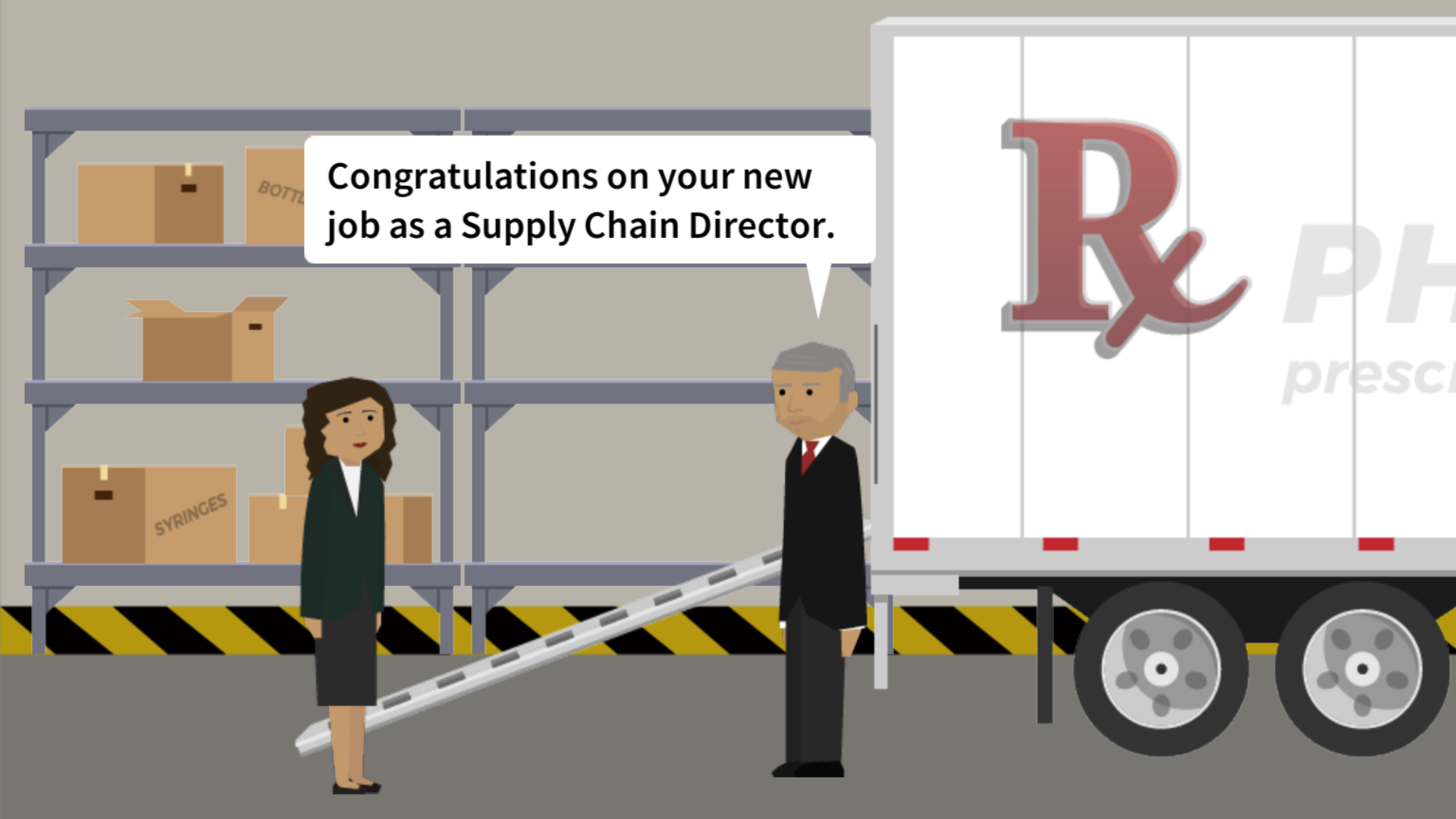}
        \caption{Briefing}\label{fig:briefing}
    \end{subfigure}
    \begin{subfigure}[t!]{.35\columnwidth}
        \centering
        \includegraphics[width=.9\columnwidth]{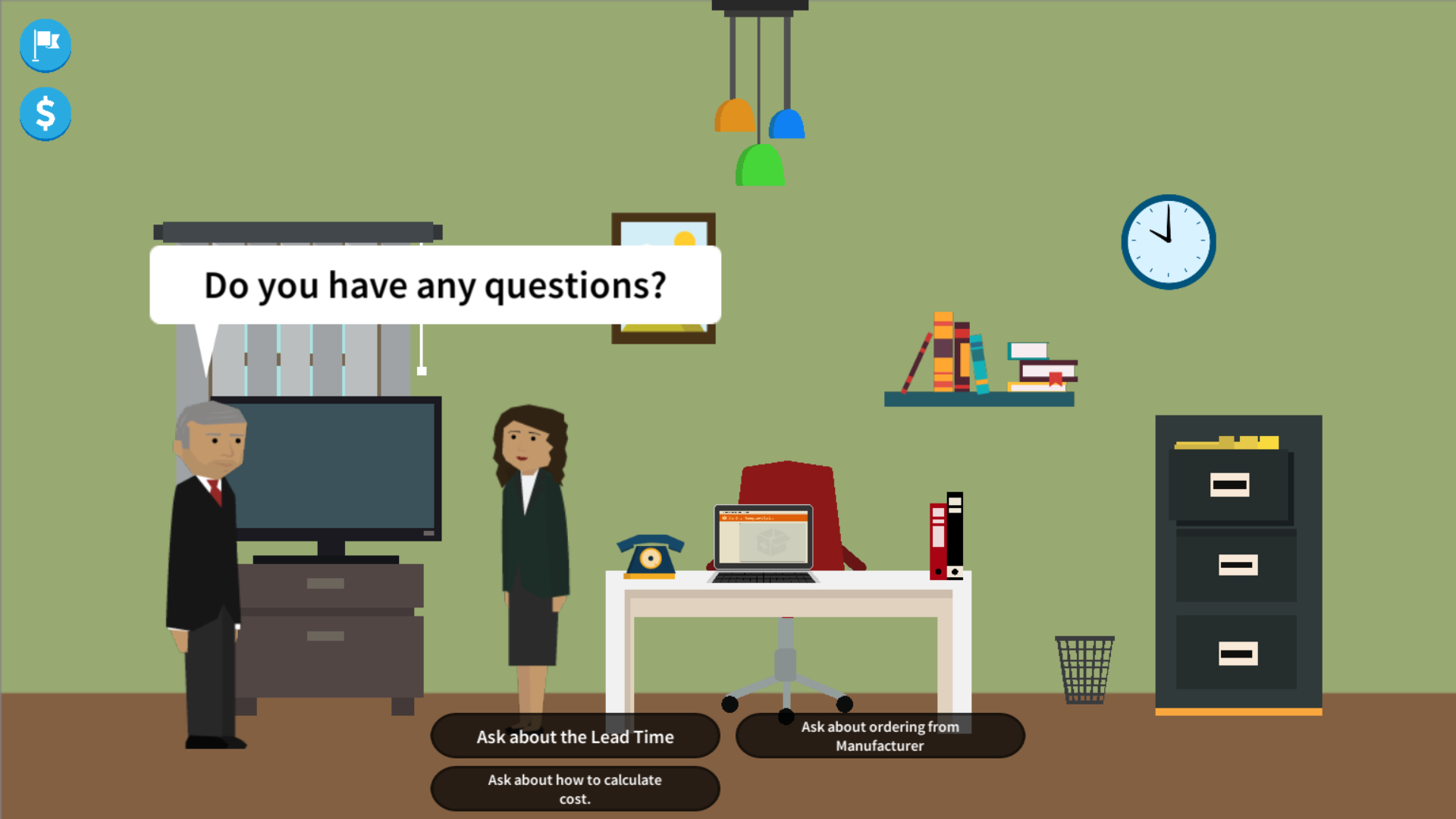}
        \caption{Tutorial}\label{fig:NPCs}   
    \end{subfigure}
    \vspace{.2cm}
    
    \begin{subfigure}[t!]{\columnwidth}
        \centering
        \includegraphics[width=.5\columnwidth]{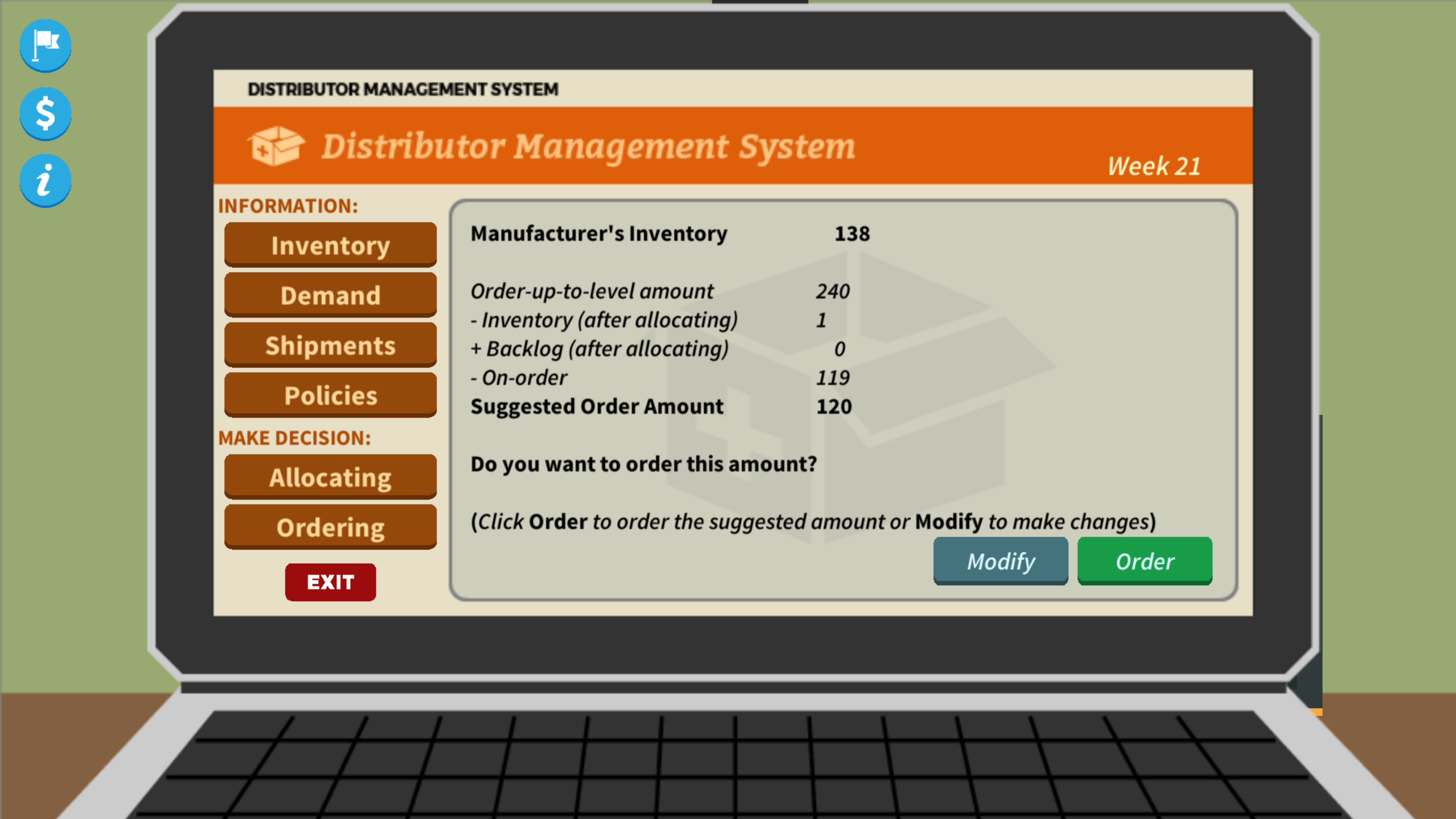}
        \caption{Gameplay}\label{fig:laptop}
    \end{subfigure}
    \caption{Gamette design: (a) Briefing player about their role and their task, (b) Tutorial through interacting with NPC, and (c) Interacting with laptop computer to make order and allocate decisions.}
    \label{fig:gamette design}
    \Description[Screenshots of gamette briefing, tutorial and gameplay phases]{The screenshot of the briefing scene shows the player character (Kate) and a Non-Player Character (Kate's boss). Players learn how to play and are informed about the study's goal through dialogue. The screenshot of the tutorial phase shows Non-Player Character (Kate's boss) explaining the role of players and their task in the game. Finally, a screenshot of the gameplay phase shows a laptop computer that players can interact with to review their inventory, demand, shipments, policies and make allocation and ordering decisions.}
\end{figure}

We created the gamette with StudyCrafter\footnote{\url{https://studycrafter.com}}, a platform where users can easily create, play, and share gamified projects. Using StudyCrafter, we designed a gamette where players take the role of a wholesaler in a drug delivery supply chain. For more details on gamette design, we refer to our previous work~\cite{mohaddesi_introducing_2020}. We considered five phases in the design of our wholesaler gamette
: (1) briefing, (2) tutorial, (3) gameplay, (4) survey, and (5) debriefing. Each game starts with a \textit{briefing} scene (see Figure~\ref{fig:briefing}) where players learn how to play and are informed about the purpose of the study. Next, the \textit{tutorial} phase starts where players see their character (Kate) and an NPC (Kate's boss) that informs them, through dialog, about their task and the goal of the game (see Figure~\ref{fig:NPCs}). The tutorial lasts for four game-weeks. During the game, players can interact with a laptop computer to observe information and make decisions. In particular, by clicking on the laptop, they will enter a management system where they can click on different buttons to observe information (inventory level, current demand,  received shipments, allocation and ordering policies) and make decisions (allocate inventory or place orders). 

After the tutorial phase, the NPC informs players that their training is over, and the \textit{gameplay} phase starts (see Figure~\ref{fig:laptop}). This phase continues for 35 weeks, and each week players can review their inventory, demand, shipments, sales revenue, and costs, then allocate their inventory and place an order. During the \textit{gameplay}, players receive feedback on their performance through interaction with other NPCs in the form of a meeting scene. This feedback includes factual information without presenting any form of bias on players' performance. We also inform players about the disruption scenario during this phase through interaction with the boss character. After Week 55, the \textit{survey} phase starts, where players are asked to answer some questions about their experience and complete a demographics survey. Finally, in the last phase, the \textit{debriefing}, players are informed about their performance, and an NPC explains the purpose of the game and experimental conditions.

\section{Methods}
\begin{figure}[ht]
  \centering
  \includegraphics[width=0.5\columnwidth]{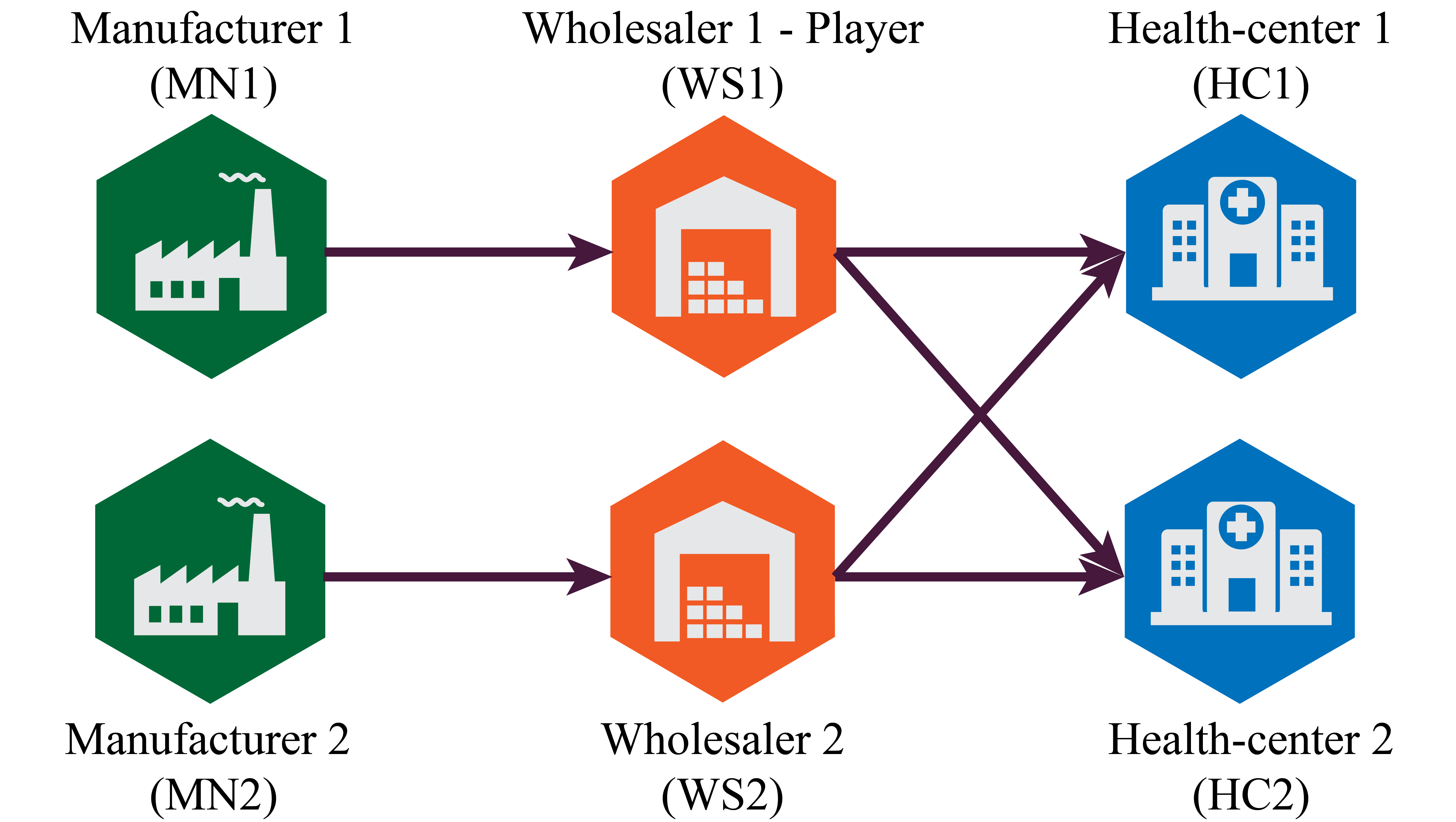}
  \caption{Supply chain network structure including two manufacturers, two wholesales and two health centers. Players play the wholesaler role using gamette.}
  \Description{The structure of supply chain network}
  \label{fig:supply_chain_network}
  \Description[Supply chain network structure]{Structure of the supply chain network, including two manufacturers, two wholesalers, and two health centers. Players play the wholesaler role using gamette.}
\end{figure}

We are interested in studying the role of human behavior in pharmaceutical supply chains. In particular, we are studying supply chain resiliency by focusing on how human decision-making affects or is affected by drug shortages. To accomplish this, we simulated a pharmaceutical supply chain network using the Flow Simulator and used a gamette to immerse human participants into a specific role within this network. We considered a supply chain network consisting of two manufacturers, two distributors, and two health centers. Figure~\ref{fig:supply_chain_network} illustrates the network structure as well as the flow of shipments between each entity in this supply chain. Compared to the network in Figure~\ref{fig:supply_chain_network}, pharmaceutical supply chains are more complex with more interdependent roles. However, this network is still complex enough to represent the behavioral dynamics of the pharmaceutical supply chains by allowing multiple agents in each echelon~\cite{doroudi2020effects}.

\subsection{Hypotheses}
While different reasons cause drug shortages, most of them can be traced back to supply chain disruptions~\cite{tucker2020incentivizing}. In the past two decades, numerous studies focused on reducing the impact of disruptions of drug shortages and increasing supply chain resiliency~\cite{tucker2020drug}, through implementing decision support systems~\cite{chihaoui2019decision} and developing optimal inventory management and ordering policies~\cite{azghandi2018minimization}. However, as described by~\cite{doroudi2020effects}, human behavior can prolong or aggravate disruptions. In our prior work~\cite{mohaddesi_introducing_2020}, we also showed that people tend to deviate from optimal order suggestions in the context of the Beer Game~\cite{sterman1989modeling}. Such behaviors can be attributed to hoarding or panic buying as an emotional response to scarcity~\cite{sterman2015m}, or the decision-maker's lack of trust in the system recommendations~\cite{wang2008attributions}, especially when the decision-maker is in the early phases of interacting with the system. As uncertainty is one of the main factors affecting trust~\cite{cho2015survey}, we expect the deviations to be higher during a shortage. Therefore, we hypothesize:\\

H1. \textit{People deviate more from the order suggestions when facing a shortage compared to a normal condition.}\\

On the other hand, the literature around drug shortages points to insufficient information sharing among different stakeholders in pharmaceutical supply chains~\cite{yang2016current}, and that we can mitigate much of the workload associated with managing shortages in such supply chains by increasing collaboration and sharing information~\cite{pauwels2015insights}. Prior research around the Beer Game showed that sharing information can reduce order fluctuations and the bullwhip effect~\cite{yang2021behavioural}. Thus, we expect that sharing information on supplier inventory level reduces the deviation of our participants from the system's order suggestions. Hence:\\

H2. \textit{Information sharing reduces the amount of deviation from order suggestions.}\\

Of course, we cannot simply assume everyone behaves similarly. For this reason, we treat participants separately and seek to characterize their behavior in answering the above hypotheses.

\subsection{Experimental Design} 
The experimental setting for our study is described in Table~\ref{tbl:experimental_design}. All human players play the role of Wholesaler 1, and all other agents are controlled via the Flow Simulator. The simulation agents make decisions based on order-up-to-level policy. According to this policy, each agent orders or produces enough product to bring their inventory position to a predefined level based on a periodic review policy with zero fixed costs~\cite{snyder2011fundamentals}. In our prior pilot study~\cite{mohaddesi2020importance}, we found initial evidence for the effect of disruption on human behavior. Hence, we are interested in studying human decision-making in the presence of disruption. In both conditions we considered Manufacturer 1 to be disrupted through a manufacturing shutdown which reduces its production capacity by 95\%. 

\begin{table}[ht]
    \caption{Summary of the experiment settings.}
    \label{tbl:experimental_design}
    \centering
    \renewcommand{\arraystretch}{1.6}
    \resizebox{0.5\columnwidth}{!}{
    \begin{tabular}{l >{\centering\arraybackslash} p{60pt} >{\centering\arraybackslash} p{60pt}}
    \toprule
    & \textbf{\renewcommand{\arraystretch}{1} Condition 1} \newline \textbf{\renewcommand{\arraystretch}{1} (No-Info)} & \textbf{\renewcommand{\arraystretch}{1} Condition 2} \newline \textbf{\renewcommand{\arraystretch}{1} (Info)} \\ 
    \cmidrule{1-3}
    Player Role & WS1 & WS1 \\
    Disrupted Manufacturer & MN1 & MN1 \\
    Information Sharing & No & Yes \\
    \bottomrule
    \end{tabular}}
\end{table}


We are also interested in testing the effect of information sharing on human decision-making, as it is one of the essential strategies to improve the resiliency of supply chains~\cite{iyengar2016medicine}. We considered two options for the level of information shared with players during the ordering process: (1) without and (2) with information sharing. Prior work on supply chains studied different types of information sharing, including downstream and upstream information sharing~\cite{yang2021behavioural}. Here, we focus on upstream information sharing. In particular, in condition 2, we inform players about their upstream manufacturer's inventory (i.e., inventory of MN1 in Figure~\ref{fig:supply_chain_network}).

Regarding the rest of our experimental design, both health-centers receive a constant demand. However, they split their orders between wholesalers differently. The HC1 agent splits its orders to its upstream agents, considering a trustworthiness measure in both conditions, meaning that HC1 orders less from the upstream wholesaler that fails to deliver drugs consistently. HC2, on the other hand, splits its orders always equally regardless of its wholesalers' trustworthiness. This trust-based behavior of HC1 particularly affects players during the disruption period (weeks 28-33), where the players who do not have enough inventory fail to satisfy HC1's demand completely. Therefore, during the disruption, in addition to incurring stockout cost for not satisfying demand, players will also experience a decrease in HC1's demand.

\subsubsection{Participants and Material}
We recruited 135 online participants through Prolific~\footnote{\url{www.prolific.co}}. We limited participation to working professionals who reported (1) English as their first language or being fluent in it; and (2) having an undergraduate degree or higher. Each participant spent on average 58 minutes (\textit{SD}=19) to play and received \$7.5 reward for their participation. Participants could access the gamette online on a web page that we created. The web page included a description of the study's purpose and a link that the gamette could be played on. Each participant was required to use a laptop or a desktop to access this web page. At the end of each gamette, we included a short survey querying participants about their experience and strategy in playing the game.

After the initial inspection of the data, we found that some participants performed extremely poorly in their game profit. Therefore, we performed outlier analysis by applying three standard deviations of game profits as the outlier threshold and observed 14 extreme outliers. The poor profit of these participants was due to incurring a high inventory cost (seven participants) as a result of ordering large amounts (i.e., an order of magnitude more than other players); high stockout cost (six participants) as a result of ordering less than the received demand for multiple periods; or both a high inventory and stockout cost (one participant) as a result of ordering less than the demand in the middle of the game and ordering large amounts towards the end. Although the specific behavior of these players is of interest, the methods we used in our analyses (i.e., PCA and HMM) are known to be sensitive to outliers. Thus, we decided to remove these players to avoid distorting our data and analyses and performed our analyses with 121 participants (47 males, 70 females, two non-binary, two preferred not to answer). The age range is 21 to 71 years (\textit{M}=34, \textit{SD}=9.9). 


\subsubsection{Incentive Design}
Previous research points to the importance of incentives in conducting experimental research~\cite{katok2018designing}. To motivate participants to engage with the task to perform well, we offered them a monetary incentive (\$50) which was gifted through a raffle. Players who performed better had a higher chance in the raffle. Each participant received one ticket for completing the game plus one ticket for every \$1000 in-game profit that they made, more than the average profit of all other players. Before starting the game, we provided each participant with instructions about how better performance could increase their chances of winning. 

\begin{figure}[ht]
    \centering
    \begin{subfigure}[t!]{.46\columnwidth}
        \centering
        \includegraphics[width=0.95\columnwidth]{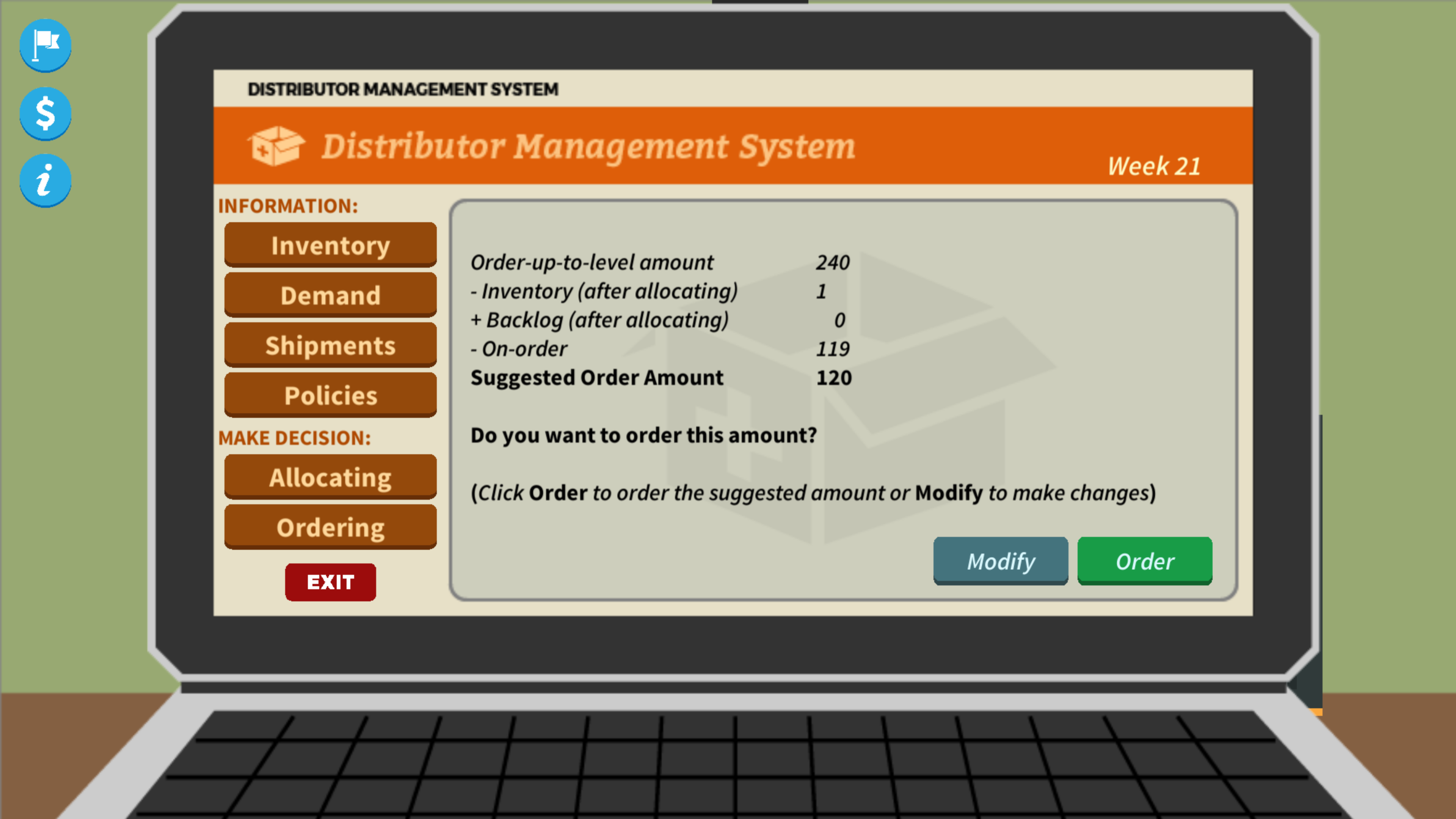}
        \caption{No information sharing on MN1 inventory }\label{fig:orderingScene1}
    \end{subfigure}
    \begin{subfigure}[t!]{.46\columnwidth}
        \centering
        \includegraphics[width=0.95\columnwidth]{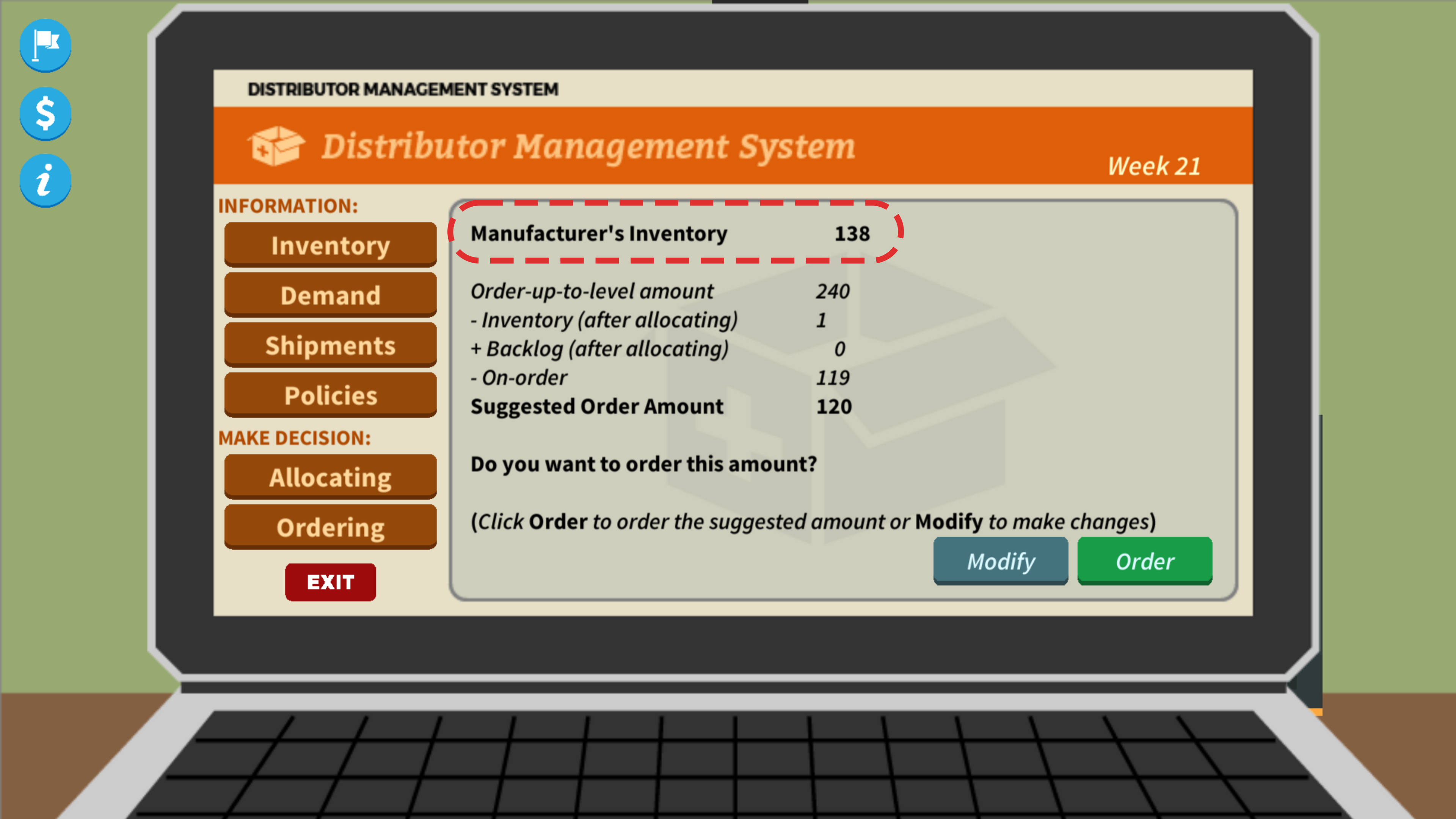}
        \caption{Sharing MN1 inventory}\label{fig:orderingScene2}   
    \end{subfigure}
    \caption{The ordering scene in each condition. Note that in (b)  "Manufacturer's Inventory" is mentioned at top while in (a) this information is not displayed.}
    \label{fig:orderingScene}
    \Description[Difference between ordering scenes in two conditions]{Ordering scenes in two conditions look identical except for the manufacturer's inventory that is provided in the information sharing condition.}
\end{figure}

\subsubsection{Procedure}
All participants first visited the study website, where they were formally briefed about the experiment and its purpose. Then, by starting the game, they were randomly assigned to one of the two conditions ($N_{No-Info}$=61, $N_{Info}$=60). The gamette in both conditions looked the same in all scenes except for the ordering scene. Participants in Condition 2 were provided with the inventory of their manufacturer, whereas the participants in Condition 1 did not receive such information (see Figure~\ref{fig:orderingScene}). Participants played the role of a character named Kate who was hired as a supply chain director in a wholesaler company. At the beginning of the game, an NPC (Kate's boss) expresses that the game's goal is maximizing the company's profit by minimizing the inventory and stockout costs and maximizing sales revenue. The NPC also informs players, through dialogue, about the sales revenue and cost breakdown (\$1 cost for each unit of inventory, \$10 cost for each unit of stockout, and \$5 revenue for each unit of sales), and the lead time of two weeks for orders (one week for orders to be processed by the manufacturer and one week for players to receive shipments).

Each participant first played four weeks of tutorial to familiarize themselves with the game (i.e., how to gather information) and was given instructions about ordering from the manufacturer and allocating to health-centers. Participants were asked to make an ordering decision at each period but only make allocation decisions when their inventory level was lower than their total demand. In cases where they had enough inventory, the game automatically allocated drugs to each health-center. After finishing the tutorial, all sales and cost data was reset, and they played the game for 35 game weeks starting at Week 21 of the simulation. The disruption started on Week 28 and ended on Week 33. We framed the disruption as a manufacturing shutdown due to COVID-19.

At each period, participants received a shipment from their upstream manufacturer and could review the inventory, demand, and backlog information. Next, if they had limited inventory compared to demand, they were asked to select one of the presented allocation policies: (1) allocate to HC1 first, (2) allocate to HC2 first, or (3) allocate proportionally. Finally, they received an order suggestion according to the order-up-to-level policy for making an ordering decision. They had the option to order the suggested amount or modify that. The gamette sends player decisions to the Flow Simulator, which moves the simulation to the next period and sends back the updated parameters for the next round to the gamette. After playing for 35 weeks, participants completed a survey. Then, through dialogue, an NPC debriefed them about the study, experimental setting, and their performance.

\subsection{Data Analysis}
For extracting behavioral patterns, explaining the interaction of players with the simulation, and showing the effect of our implemented stimuli (i.e., information sharing), we attempt to abstract the decision space by first characterizing the system's states and then characterizing players' responses. Our goal is to find common response patterns in groups of people. Next, we describe our approach for characterizing the system and the players' responses.

\subsubsection{Characterizing the System}
For each agent within the Flow Simulator, the state of the system at each time step $t$ is described by a set of supply chain parameters $(Inv_{t},  Dem^{i}_{t}, Blg_{t}, Shp_{t}, Oor_{t})$ where:

\begin{itemize}
    \item $Inv_{t}$ is the inventory level of the agent at time $t$ after receiving shipments from upstream supplier (or from production for a manufacturer agent) and satisfying all downstream demand. 
    \item $Dem^{i}_{t}$ is the demand from downstream agent $i$ at time $t$.
    \item $Blg_{t}$ is the total backlog amount (i.e. unsatisfied downstream demand) by time $t$.
    \item $Shp_{t}$ is the received shipment/production amount at time $t$.
    \item $Oor_{t}$ is total on-order amount which is the amount that has been ordered from the upstream supplier by time $t-1$ and has not been received. For a manufacturer agent this is equivalent to the in-production amount.
\end{itemize}

We used Principal Component Analysis (PCA) to reduce the dimensionality of the state space for the players' role (i.e., the wholesaler). We did this by first standardizing the parameters mentioned above for all players and across all periods. For applying PCA, we used the scikit-learn in Python~\cite{scikit-learn}. Then, we performed hierarchical cluster analysis~\cite{johnson1967hierarchical} on the scores from the emerged components to characterize the states of the system.

\subsubsection{Characterizing the Behavior}
Hidden Markov Models (HMMs) are useful for characterizing sequential patterns~\cite{rabiner1989tutorial} and particularly for behavioral modeling of game players~\cite{hooshyar2018data}. An HMM is defined by a set of parameters $(S_{t}, O_{t}, A, B, \pi)$ where:

\begin{itemize}
    \item $S_{t}$ is the finite set of hidden states.
    \item $O_{t}$ represents the finite set of observed outputs.
    \item $A$ is the transition probability matrix that indicates the probability of moving from one state to another.
    \item $B$ is the emission matrix (i.e., observation probability matrix) that indicates the probability of seeing each observation in each state.
    \item $\pi$ is the initial state probability matrix which represents the probability of being in a given state at the start of the sequence.
\end{itemize}

\begin{figure*}[ht]
    \centering
    \includegraphics[width=\textwidth]{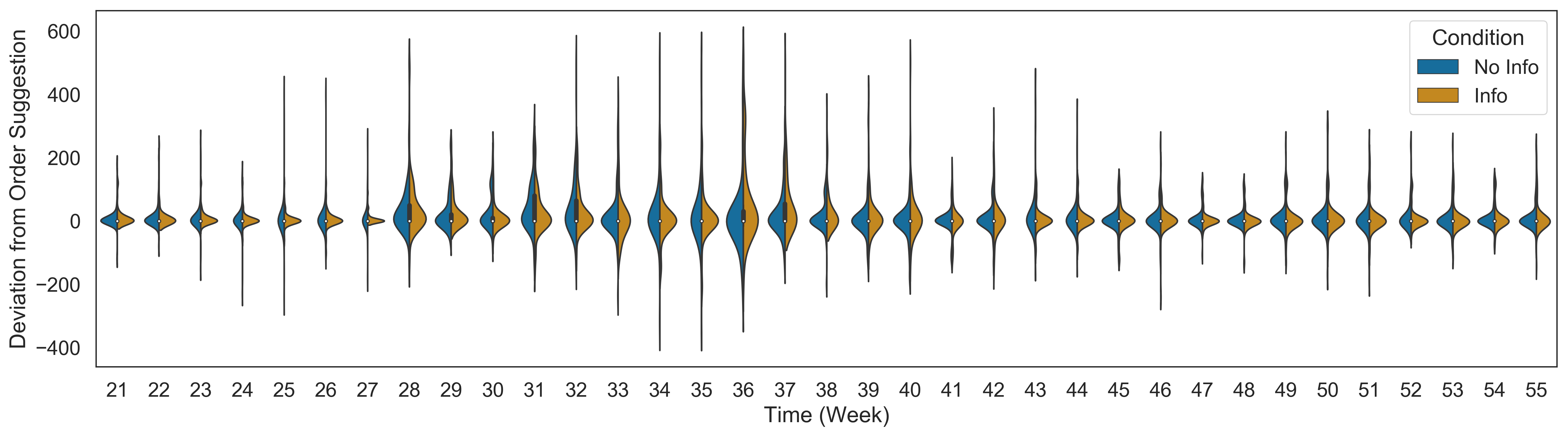}
    \captionsetup{width=\textwidth}
    \caption{Distribution of players' deviation from suggested order amounts over time. At each period, the left density curve shows the distribution of order deviations for players in the No-Info condition and the right density curve shows the same for players in the Info condition.}\label{fig:hmm_observations}
    \Description[Distribution of order deviations for the players in each condition]{The plot shows the distribution of players’ order deviations in each period during the game in the form of density curves. The left density curve indicates the distribution of order deviations for players in the No-Info condition at each period. The right density curves show the same for players in the Info condition. The distribution of deviations shows a more considerable variance during the shortage period, which is after Week 28 for both conditions.}
\end{figure*}

We used HMM for characterizing the players' behavior in their sequential decisions during their gameplay. In particular, we considered players' order deviations from the suggested order amounts as the observed outputs ($O_{t}$) of HMM. Figure~\ref{fig:hmm_observations} shows the distribution of players' order deviations in the form of density curves in each period during the game. Our goal is to find the optimal sequence of hidden states that can best explain the observed decisions by the players, where the hidden states ($S_{t}$) represent the players' response mode at time $t$. 

\begin{figure}[ht]
    \centering
    \includegraphics[width=.5\columnwidth]{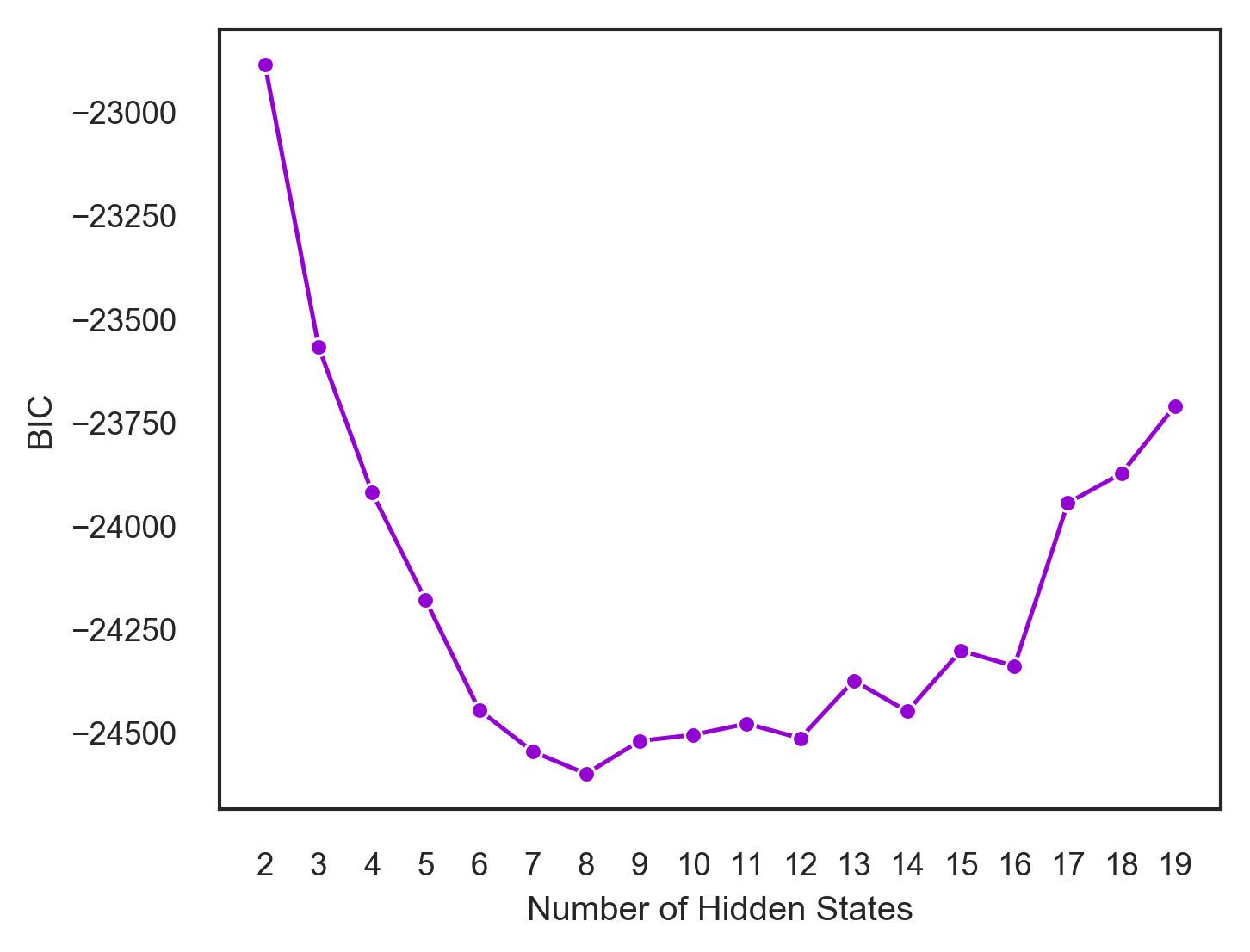}
    \captionsetup{width=\linewidth}
    \caption{BIC score for different number of hidden states. The plot suggests that considering eight hidden states maximizes the likelihood of the model for our observed data.}\label{fig:bic_scores}
    \Description[Bayesian Information Criterion score for different number of hidden states]{The plot shows the Bayesian Information Criterion scores for running the Baum-Welch algorithm on observed data for different numbers of hidden states. The plot indicates that having eight hidden states minimizes the Bayesian Information Criterion scores and maximizes the likelihood of the model for the observed data.}
\end{figure}

We adopted the approach used by prior researchers for learning the model parameters and finding the optimal sequence of hidden states~\cite{bunian2017modeling, carola2011hidden}. According to this approach, we first used an implementation of the Expectation-Maximization (EM) algorithm, known as the Baum-Welch algorithm~\cite{bilmes1998gentle}, in Python to learn the HMM parameters from the players' data. The algorithm iteratively re-estimates the model parameters to maximize the likelihood that the HMM generates the observed data. Next, using the estimated model parameters, we used an implementation of the Viterbi algorithm~\cite{forney1973viterbi} in Python to obtain the sequence of hidden states that best explain each player's decision. The obtained sequence is a list of labels representing the response modes of players at each period that led to a specific decision. To determine what number of hidden states ($S$) is optimal, we ran the Baum-Welch algorithm multiple times with different numbers of states. Then, we calculated the Bayesian Information Criterion (BIC) to identify what number of hidden states minimizes the BIC and maximizes the model likelihood. BIC has shown to be an effective measure for calculating the number of states in HMM, especially when the observed data are independent~\cite{celeux2008selecting}. We calculated the BIC using the following equation:

\begin{equation}
    BIC = -2\log Likelihood + p \log(D)
\end{equation}

where the $Likelihood$ is the likelihood of the model given observed data, $p$ is the number of free parameters in the model, and $D$ is proportional to the size of the data. The number of free parameters is the sum of the free parameters for each model component (i.e., transition probabilities, emission matrix, and initial state probabilities)~\cite{li2017incorporating}. Figure~\ref{fig:bic_scores} illustrates BIC scores for running the Baum-Welch algorithm on observed data for various numbers of hidden states. BIC scores suggest that considering eight hidden states maximizes the likelihood of the model for the observed players' decisions.

\subsubsection{Analyzing Survey Responses}
In our survey at the end of each game, we asked players about their experience (``Can you elaborate on how you experienced playing this game?'') and strategy (``What was your strategy in playing this game?'') in playing the game. Using Initial and Axial Coding~\cite{saldana2021coding}, we qualitatively coded the open-response comments from players. This analysis served to triangulate the results from the aforementioned analyses. In particular, we first coded players' responses independently from other analyses. We then triangulated the results by linking the outcomes of this coding to the player types we identified through HMM. We performed this process separately for the experience question and the strategy question. We use the results from the strategy question to describe player types in Section~\ref{behavior_char}. As for the experience question, while we did not find much insight from the triangulation, we identified how different players experienced the game, which we discuss in Section~\ref{reinstitue_hsi}. 

We followed an inductive approach in our coding practice, where we used a combination of in vivo and constructed codes. After familiarizing ourselves with the data, one researcher coded the comments and generated initial categories. Then, another researcher reflected on the codes and reviewed the generated categories. Finally, we triangulated the generated categories across identified player types (i.e., hoarders, reactors, followers). Players' comments and generated categories are available in the supplementary materials. In Section 5, we refer to participant quotes as "(Player ID, gender, age)". For example, a female participant with age 22 and player ID 44 would be displayed as "(PL44, female, 22)".

\section{Results}
\subsection{Characterizing the System}\label{system_char}
\begin{figure}[ht]
    \centering
    \includegraphics[width=.5\columnwidth]{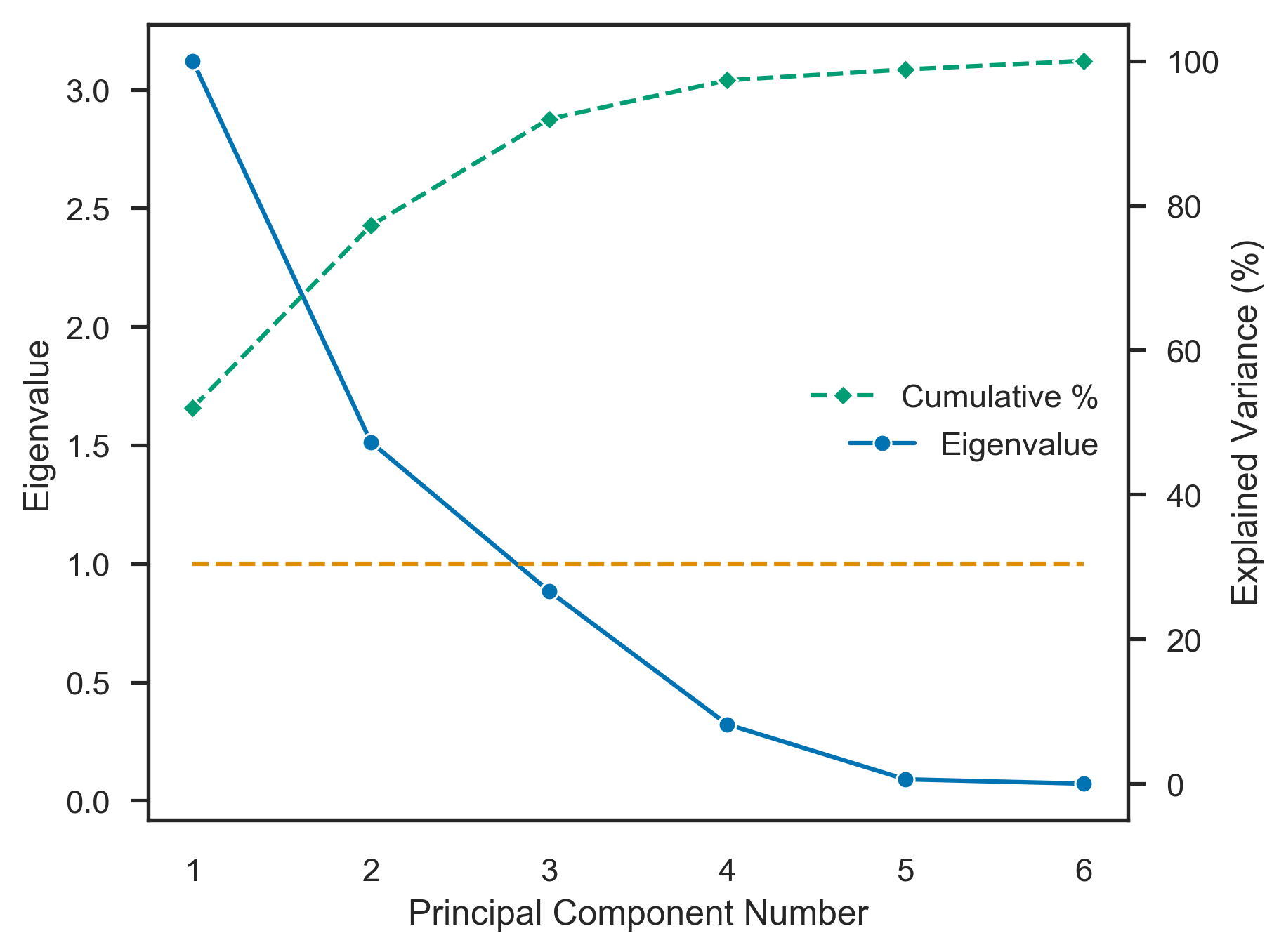}
    \captionsetup{width=\linewidth}
    \caption{Scree plot of eigenvalues and explained variance of principal components. The plot shows the first and second principal components explain 77\% of the variance in data.}\label{fig:scree_plot}
    \Description[Scree plot of eigenvalues and explained variance of each principal component]{Scree plot of eigenvalues and explained variance of each principal component which shows the first and second principal components explain 77\% of the variance in data.}
\end{figure}

We first applied Principal Component Analysis (PCA) to reduce the dimensionality of the state space including \textit{inventory} ($Inv$), \textit{demand of HC1} ($Dem^{HC1}$), \textit{demand of HC2} ($Dem^{HC2}$), \textit{backlog} ($Blg$), \textit{received shipment} ($Shp$), and \textit{on-order} ($Oor$) for all players and across the 35 game-weeks they played. A scree plot of the eigenvalues (see Figure~\ref{fig:scree_plot}) suggested two components explain 77\% of the variance in data with the first component explaining 52\% and the second component 25\% of the variance. Table~\ref{tbl:pca_correlations} shows the correlations of the principal components with each supply chain parameter. Component 1 shows positive correlation with \textit{backlog} and \textit{on-order} and negative correlation with \textit{demand} from both health-centers. Therefore, an increase in Component 1 scores is indicative of a supply problem where players cannot satisfy demand, hence, face an increase in their \textit{backlog} and \textit{on-order}, and decrease in their customers' \textit{demand}. Thus, Component 1 can be considered as a measure of \textit{supply-side disruption}. Component 2, on the other hand, shows positive correlation with \textit{inventory} and \textit{received shipment}, suggesting an increase in Component 2 scores is indicative of an increase in the level of \textit{inventory} and \textit{received shipment}. Therefore, we consider Component 2 to be a measure of \textit{recovery after disruption} where participants start to receive large shipments after a period of shortage.

\begin{table}[ht]
    \caption{Pearson correlation coefficients between principal component scores and each supply chain parameter.}
    \label{tbl:pca_correlations}
    \centering
    \renewcommand{\arraystretch}{1.7}
    \resizebox{0.5\columnwidth}{!}{
        \begin{tabular}{l c c c c c c}
        \toprule
        & $Inv$ & $Dem^{HC1}$ & $Dem^{HC2}$ & $Blg$ & $Shp$ & $Oor$ \\ 
        \cmidrule{1-7}
        \textbf{Component 1} & -0.11 & \textbf{-0.88} & \textbf{-0.9} & \textbf{0.89} & -0.05 & \textbf{0.86} \\
        \textbf{Component 2} & \textbf{0.51} & -0.38 & -0.33 & -0.37 & \textbf{0.9} & -0.24 \\
        \bottomrule
        \multicolumn{7}{m{9cm}}{\small Note: Correlations greater than 0.4 or less than -0.4 are in bold. All correlations are significant at $\alpha = 0.01$.}
    \end{tabular}
    }
\end{table}

\begin{figure}[ht]
    \centering
    \includegraphics[width=.5\columnwidth]{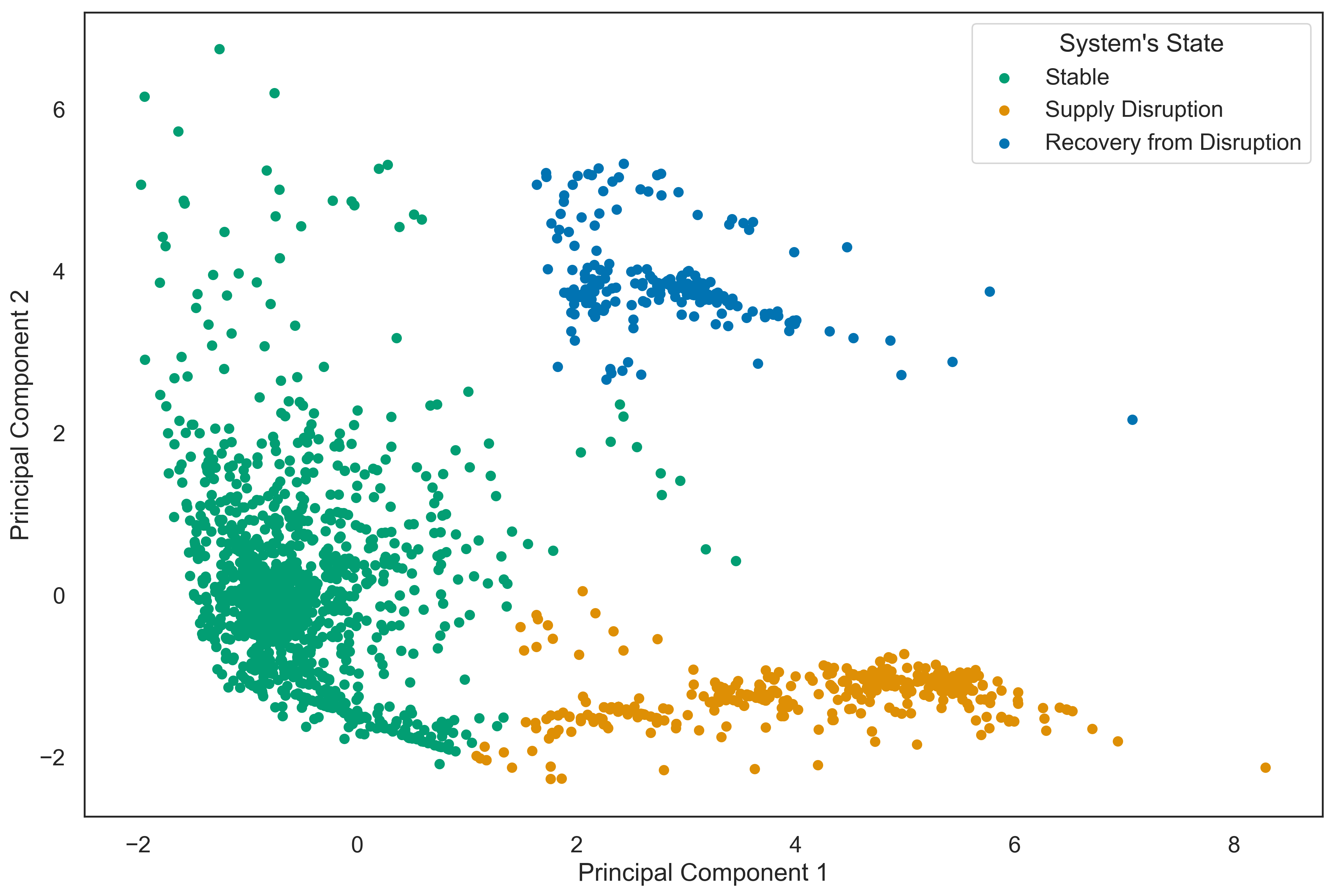}
    \captionsetup{width=\linewidth}
    \caption{Principal components and their scores. Increase in Principal Component 1 scores correlates with increase in backlog and on-orders representing a state of supply disruption. Increase in Principal Component 2 scores correlates with increase in inventory and received shipments representing recovering from disruption.}\label{fig:pca_scores}
    \Description[Principal component scores and the emerged clusters]{The scatter plot shows principal component scores and clusters, which refer to specific system states that players experienced, including stable, supply disruption, and recovery from disruption. An increase in Principal Component 1 scores correlates with the increase in backlog and on-orders, representing a state of supply disruption. An increase in Principal Component 2 scores correlates with an increase in inventory and received shipments representing recovering from disruption.}
\end{figure}

Next, we performed hierarchical cluster analysis using the scores from the two principal components, where three distinct clusters emerged. Figure~\ref{fig:pca_scores} displays component scores and the emerged clusters. These clusters represent specific states of the system that players experienced. We call these states \textit{stable}, \textit{supply disruption}, and \textit{recovery from disruption}. The \textit{stable} state represents the situation where players consistently receive shipments from their supplier and satisfy the demand from health centers. As can be observed in Figure~\ref{fig:pca_scores}, \textit{supply disruption} state corresponds to an increase in the scores of Component 1. Therefore, in this state, players experience an increase in their \textit{backlog} and \textit{on-order} meaning they do not receive enough shipments from MN1 to satisfy their demand. An increase in Component 1 scores also indicates a decrease in demand from both health centers. It is intuitive as HC1 holds a trustworthiness measure and orders less from players when they fail to deliver. The HC2 agent also slightly reduces its orders from players since it follows the order-up-to-level policy. Finally, the \textit{recovery from disruption} state corresponds to an increase in Component 2 scores, suggesting an increase in \textit{received shipments} and \textit{inventory}.

\begin{figure}[ht]
    \centering
    \includegraphics[width=0.7\columnwidth]{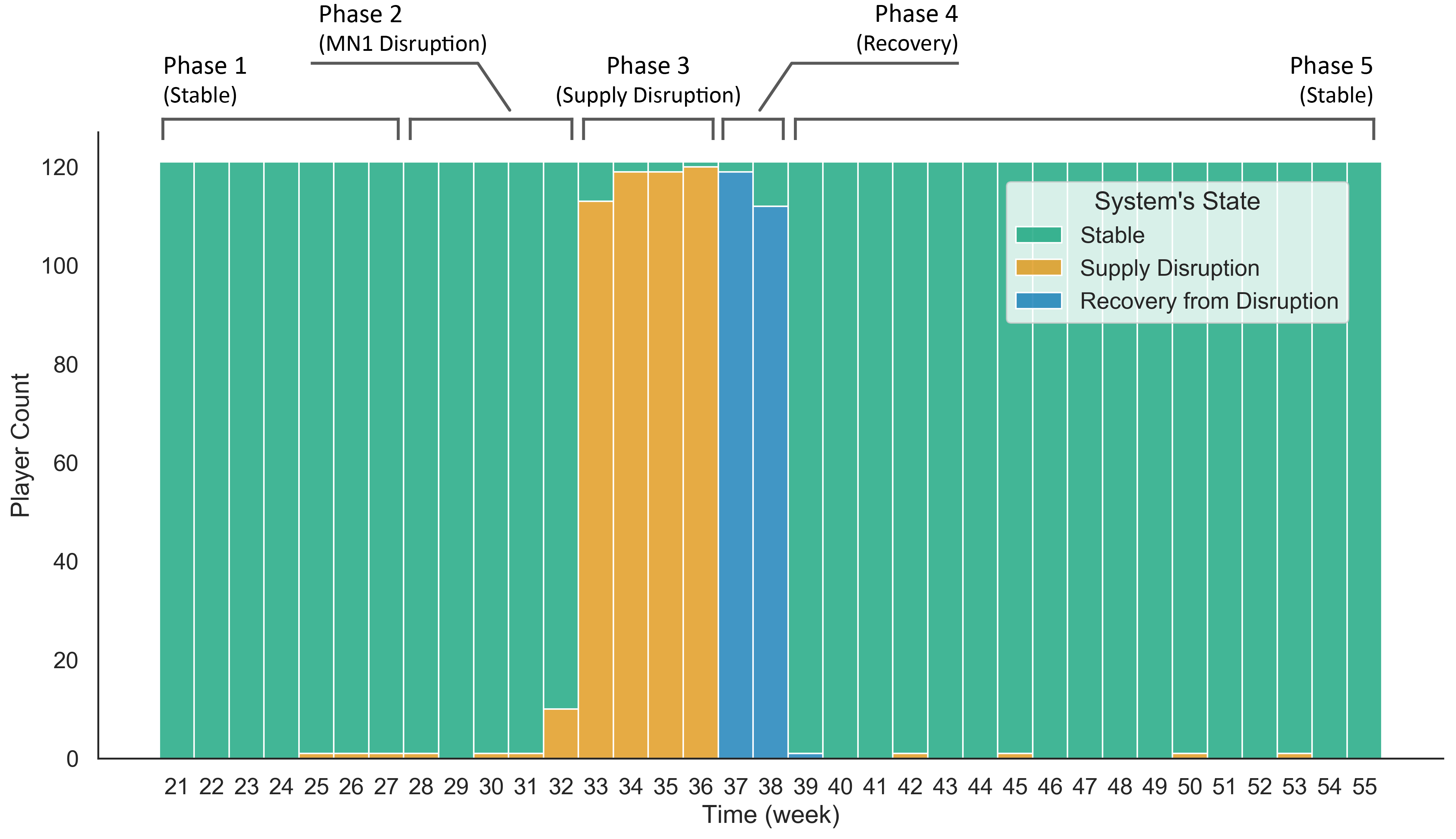}
    \captionsetup{width=\linewidth}
    \caption{Count of players in each system's states and the corresponding system phases based on PCA cluster analysis. Each phase represents a specific state. Note that Phase 2 starts with Week 28 which is when MN1 is disrupted.}\label{fig:system_states}
    \Description[System phases resulting from the players' counts in each state]{Stacked barplot illustrating the count of players in each system state and the corresponding five system phases.}
\end{figure}

To characterize the systems' states temporally, we counted the number of players in each state over time. We plotted these counts in Figure~\ref{fig:system_states}. According to this plot, the majority of players experience a \textit{stable} system state until Week 32. Then, the players start to experience the impact of disruption in Manufacturer 1 (i.e., the \textit{supply disruption} state) starting Week 33 and until Week 36. At Week 37 and 38, the system transits to the \textit{recovery from disruption} state, where the majority of players start to receive large shipments from the recovered manufacturer. Finally, with the start of Week 39, the system transits back to the \textit{stable} state for the majority of players. We also noticed that the described pattern does not hold for a few players. For example, there are a few players who experienced a \textit{supply disruption} state before Week 33 or after Week 39. In addition, the system state for player \textit{PL77} was \textit{stable} throughout the whole game.~\footnote{These behaviors are attributed to the specific decisions of players, which can be replicated. For example, \textit{PL77} over-ordered from the beginning and for every week leading up to disruption, and as a result, had enough inventory to avoid any backlog during the disruption period.} The reason for such patterns is that players' behavior also affects the state of the system. It is intuitive considering that the underlying system is a dynamic simulation where individuals can experience a unique trajectory depending on their behavior. However, the impact of the manufacturing disruption is so strong that the majority of players experience the \textit{supply disruption} and \textit{recovery from disruption} states. Therefore, we base our system characterization on the states realized by the majority of the players.

We considered five phases in which the system is in one of the described states (see Figure~\ref{fig:system_states}). We will use these five phases to characterize the interaction of players with the simulation. We intentionally divided Weeks 21-32 into two phases because at Week 28 we inform the players about the manufacturing disruption. We hypothesize that this information affects the players' interaction even though they are still in the \textit{stable} state. Next, we describe our characterization of the players' behavioral responses.

\subsection{Characterizing the Behavior}\label{behavior_char}
\begin{figure}[ht]
    \centering
    \includegraphics[width=.5\columnwidth]{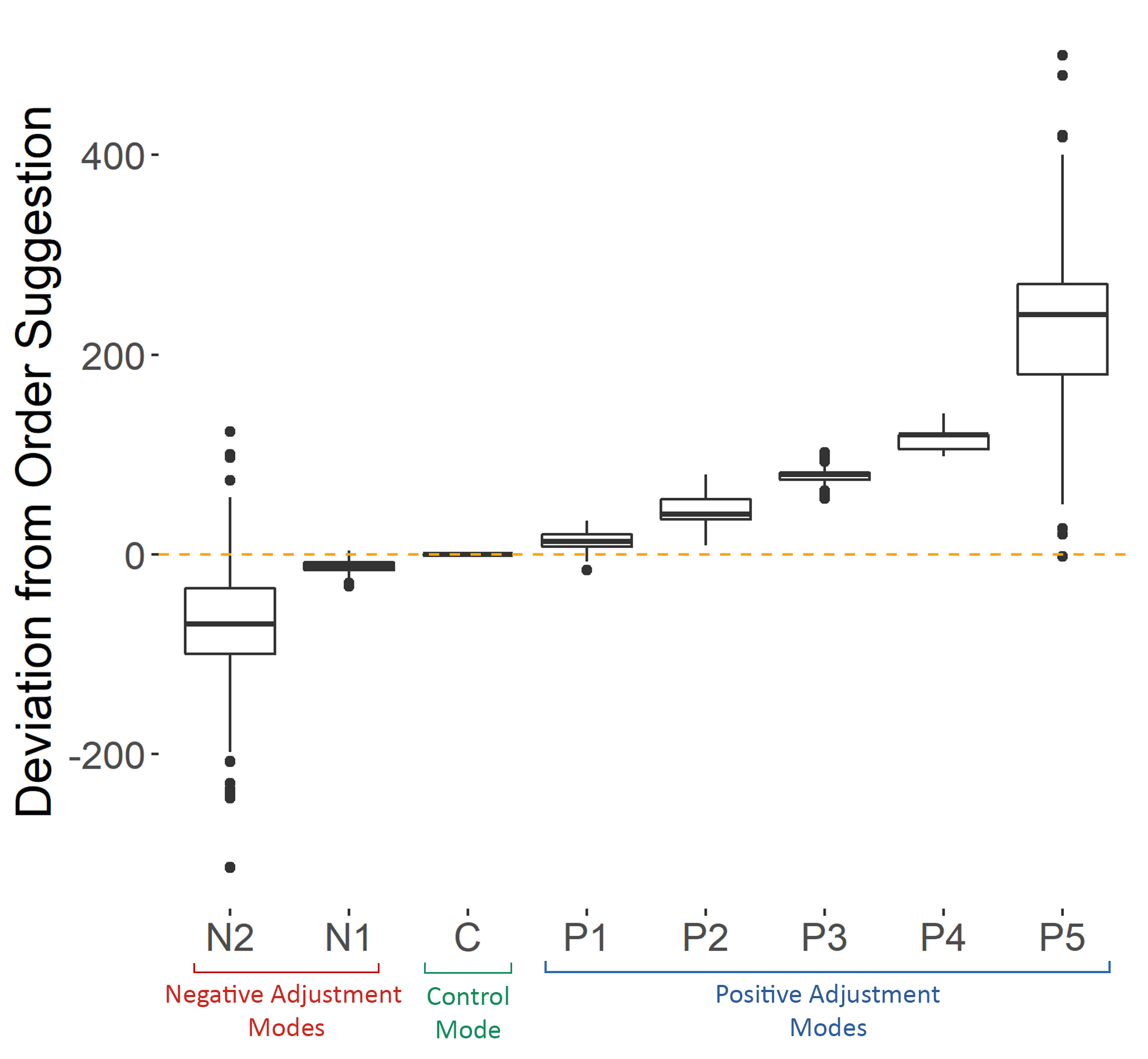}
    \captionsetup{width=\linewidth}
    \caption{HMM emission distributions for deviations from order suggestions in each hidden state (i.e., response mode). Each response mode corresponds to a certain level deviation from order suggestions.}\label{fig:hmm_emissions}
    \Description[HMM emission distributions each hidden state]{HMM emission distributions for deviations from order suggestions in each hidden state representing a behavioral response mode. Each response mode corresponds to a certain level of deviation from order suggestions in the form of boxplots. The boxplot communicates the mean, interquartile range, standard deviations, and outlier values for deviations from order suggestions for each response mode.}
\end{figure}

\begin{figure}[ht]
    \centering
    \includegraphics[width=0.5\columnwidth]{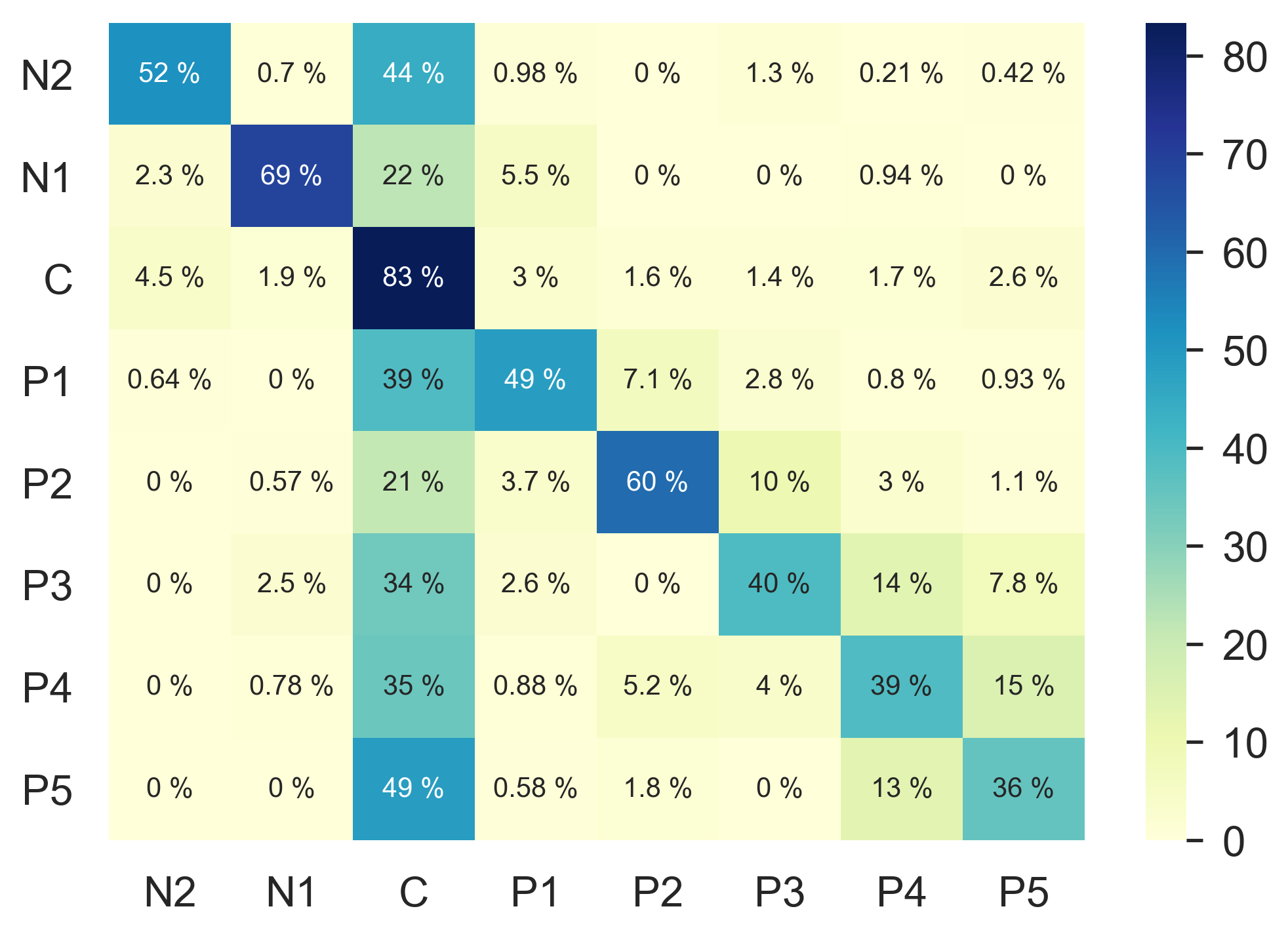}
    \captionsetup{width=\linewidth}
    \caption{Transition rates of the HMM hidden states (i.e., response modes) for all players. Each mode corresponds to a certain level of deviation from order suggestions.}\label{fig:hmm_transition_rates}
    \Description[Transition rates of response modes derived from HMM]{Transition rates of response modes derived from HMM that shows the probability of moving from one state to another. Players stay in each response mode or move to the control mode with a higher probability.}
\end{figure}

We modeled an HMM with eight hidden states using the normalized deviations from the suggested order amounts as the observation sequences, where each hidden state represents a behavioral response mode. According to this model, we found two modes of negative adjustment (N1 and N2), a control mode (C), and five modes of positive adjustment (P1, P2, P3, P4, P5). Figure~\ref{fig:hmm_emissions} shows the emission distributions for each response mode, and Figure~\ref{fig:hmm_transition_rates} illustrates the transition rates of moving from one response mode to another. Next, we obtained the sequence of response modes that best described each player's decisions by running the Viterbi algorithm over the observed data. We used the obtained sequences of response modes across all players to characterize the behavior.

\begin{table}[h]
	\centering
	\caption{Contingency table of frequency of players who contained each response mode in their sequences.}
	\label{tab:contingencyTable}
	\resizebox{.7\columnwidth}{!}{
	\begin{threeparttable}
		\begin{tabular}{lrrrrrrrrrr}
			\toprule
			\multicolumn{1}{c}{} & \multicolumn{1}{c}{} & \multicolumn{8}{c}{Response Mode} & \multicolumn{1}{c}{} \\
			\cline{3-10}
			Player Type &  & N2 & N1 & C & P1 & P2 & P3 & P4 & P5 & Total  \\
			\cmidrule[0.4pt]{1-11}
			Hoarder & Count & 28 & 24 & 55 & 43\tnote{$\ast$} & 32 & 41\tnote{$\ast\ast$} & 34 & 32 & 289  \\
			($n=56$) & Expected & 34.4 & 24.7 & 68.9 & 39 & 29.3 & 31 & 31.6 & 29.8 & 289  \\
			Reactor & Count & 30 & 19 & 48 & 25 & 19 & 13 & 20 & 20 & 194  \\
			($n=48$) & Expected & 23.1 & 16.5 & 46.2 & 26.2 & 19.6 & 20.8 & 21.2 & 20 & 194  \\
			 Follower & Count & 2\tnote{$\ast$} & 0\tnote{$\ast$} & 17 & 0\tnote{$\ast$} & 0\tnote{$\ast$} & 0\tnote{$\ast$} & 1\tnote{$\ast$} & 0\tnote{$\ast$} & 20  \\
			($n=17$) & Expected & 2.4 & 1.7 & 4.7 & 2.7 & 2.02 & 2.1 & 2.1 & 2 & 20  \\
			Total & Count & 60 & 43 & 120 & 68 & 51 & 54 & 55 & 52 & 503  \\
			($n=121$) & Expected & 60 & 43 & 120 & 68 & 51 & 54 & 55 & 52 & 503  \\
			\bottomrule
		\end{tabular}
		\begin{tablenotes}
            \small
            \item Counts show the presence of each response mode in sequences of players.
            \item[$\ast$] Chi-square test of independence shows significant association between player type and response mode at $\alpha=0.05$.
            \item[$\ast\ast$] Chi-square test of independence shows significant association between player type and response mode at $\alpha=0.01$.
        \end{tablenotes}
    \end{threeparttable}
	}
\end{table}

\begin{figure*}[ht]
    \centering
    \includegraphics[width=\textwidth]{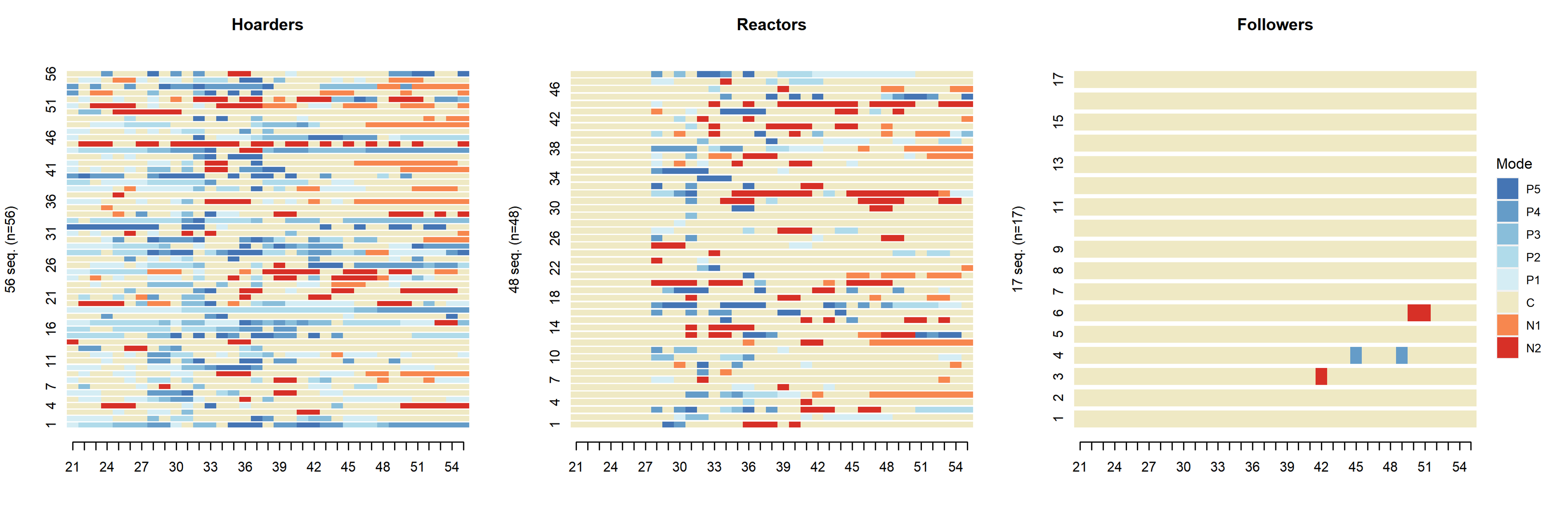}
    \captionsetup{width=\textwidth}
    \caption{Sequences of response modes for the three players types. Each row illustrates the sequences of response modes for one player. Darker shades correspond to higher level of deviation.}\label{fig:hmm_player_types}
    \Description[Sequences of response modes for the three player types]{Sequences of response mode for the three players types, including hoarders, reactors and followers. Each row illustrates the sequences of response modes for one player where shades of blue represent positive adjustment and shades of red represent negative adjustment. Darker shades correspond to higher level of deviation.}
\end{figure*}

We analyzed all players' sequences of response modes by computing the similarities between their sequences. Our goal was to find similar patterns in the sequences of response modes to classify players into different types. We used the length of the longest common prefix (LCP) proposed by~\cite{elzinga2006sequence} as the similarity measure and applied agglomerative hierarchical clustering~\cite{kaufman2008agglomerative} on the obtained similarity matrix. We found three clusters of players based on their sequences of response modes (see Appendix~\ref{appendix:optimal_clusters} for more details). We compared their response mode patterns and frequency of response modes in the sequences of each group to characterize the behavior of each type of player. Below we provide details of our characterization where we call these players: (1) hoarders (i.e., players who stockpile more frequently), (2) reactors (i.e., players who follow the systems' order suggestions before the disruption but react to the news about the disruption and transit to other response modes), and (3) followers (i.e., players who almost always follow order suggestions). Figure~\ref{fig:hmm_player_types} displays the response mode sequences, and Figure~\ref{fig:hmm_transitions_for_player_types} shows the transition rates between response modes for each player type. We also qualitatively coded the players' responses in our survey, asking about their playing strategy, and compared the codes between each type of player. In what follows, we characterize the behavior of each type of player by comparing their sequences of response modes and connecting them with players' open responses.

\subsubsection{Hoarders}
\begin{figure*}[ht]
    \centering
    \includegraphics[width=\textwidth]{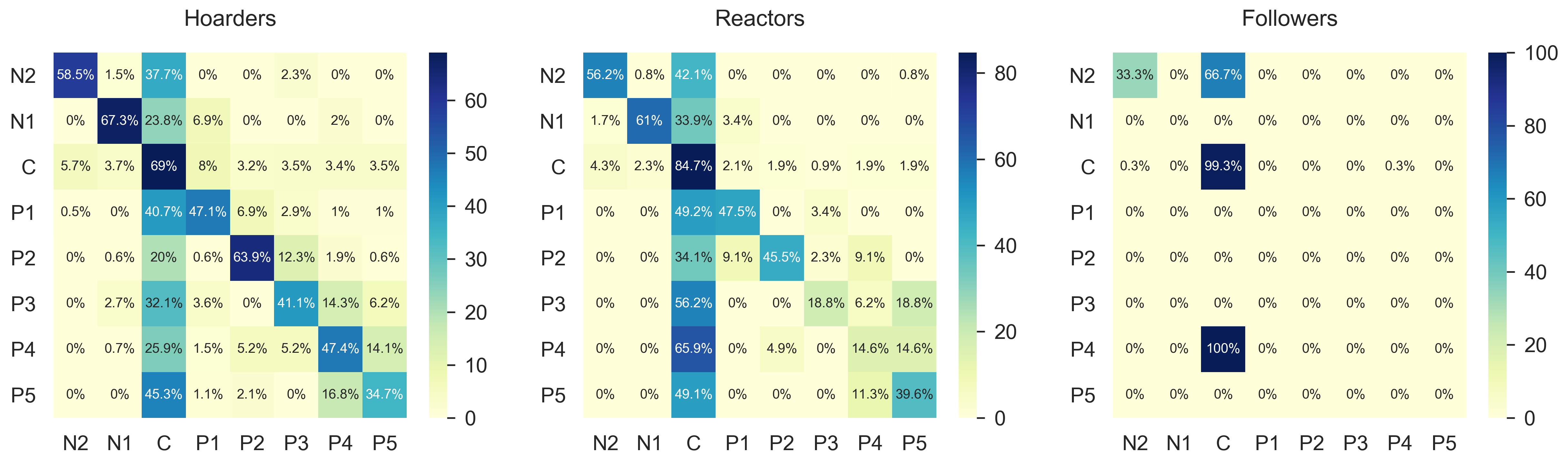}
    \captionsetup{width=\textwidth}
    \caption{Transition rates between response modes for the three player types. Followers stay in control mode with a high probability. Hoarders stay in or transit to other positive adjustment modes with a higher probability compared to reactors.}\label{fig:hmm_transitions_for_player_types}
    \Description[Transition rates between response modes for the three player types]{Transition rates between response modes for the three player types of hoarders, reactors, and followers where the rates represent probabilities. Followers stay in control mode with a high probability. Hoarders stay in or transit to other positive adjustment modes with a higher probability compared to reactors.}
\end{figure*}

We call the first group hoarders as they over-order more frequently. Hoarders include 46\% of the players ($n=56$). Half of the hoarders start the game in a positive adjustment mode ($n_{P1}=19$, $n_{P2}=5$, $n_{P4}=3$, $n_{P5}=1$), and the rest in control mode ($n_{C}=27$) and negative adjustment mode ($n_{N2}=1$). According to transition rates (see the left transition matrix in Figure~\ref{fig:hmm_transitions_for_player_types}), in general, hoarders stay in or transit to other positive adjustment modes with a higher probability compared to reactors. The analysis of response modes' presence in player sequences also suggests that hoarders tend to be more frequently in positive adjustment mode of P1 ($p=.048$) and P3 ($p=.002$) compared to reactors and followers (see Table~\ref{tab:contingencyTable}). This behavior is consistent with our qualitative data analysis, suggesting that hoarders tend to over-order. Although the strategy of some players in this group was to ``follow suggestions'' ($n=7$), many others expressed that their strategy was to ``stockpile'' ($n=22$), ``keep safety stock'' ($n=7$), ``ignore suggestions'' ($n=2$), or ``adjust suggestions'' ($n=2$), as exemplified by the quotes below: 

\begin{quote}
    \textit{``My strategy was to stock up on the supplies for at least an extra month in anticipation of supply chain issues. The saline does not cost much compared to the revenue''}--(PL87, female, 27) \\
    \textit{``Have a 2 weeks supply of spare inventory and order 10\% more than the current request volume. Ignor [sic] the recommended order amounts.''}--(PL95, male, 33)
\end{quote}

The reason for such behavior can be attributed to players' effort to ``avoid backlog'' ($n=7$) or to prepare for ``uncertainties'' ($n=2$):

\begin{quote}
    \textit{``You are punished [sic] more for backlogs than for having excess stock on hand so I tried to have extra stock to avoid the punishment [sic] of backlogs.''}--(PL6, male, 30) \\
    \textit{``..I tried to predict what would happen in the next couple of weeks as the manufacturer sometimes goes down...''}--(PL54, female, 35)
\end{quote}

Only a few players indicated they were ``experimenting'' ($n=2$) to learn how the game works: \textit{``At first, I was just trying to order and get the hang of it.''}--(PL93, female, 35).

\subsubsection{Reactors}
Reactors follow order suggestions, but as their name suggests, react to the disruption. Reactors include 40\% of the players ($n=48$). They all start the game in control mode (C) and stay in control mode until Week 28 where the NPC character informs them about the manufacturing disruption. Some of these players ($n=22$) react to the news about the disruption and transit to a different response mode ($n_{N2}=3$, $n_{N1}=1$, $n_{P1}=6$, $n_{P2}=2$, $n_{P3}=1$, $n_{P4}=8$, $n_{P5}=1$) instantly on Week 28 (see Figure~\ref{fig:hmm_player_types}). The rest ($n=26$) show this reaction at a later time during the disruption period. The frequencies in Table~\ref{tab:contingencyTable} do not show any response mode to be significantly more or less frequent across reactors. However, compared to hoarders, reactors stay in positive adjustment modes (specifically P2, P3, P4) with less probability and move back to the control mode with a higher probability (see the middle transition matrix in Figure~\ref{fig:hmm_transitions_for_player_types}); indicating reactors tend to follow order suggestions with a higher probability. Our qualitative data analysis shows similar results. Reactors' strategy was to ``follow suggestions'' ($n=15$):

\begin{quote}
    \textit{``My strategy was the follow the sugesstions [sic].''}--(PL18, female, 25) \\
    \textit{``..eventually I realized ordering the suggested amount was getting me a higher profit so I kept with that.''}--(PL25, female, 27) \\
    \textit{``Looking at what we had on hand and what the trendsw [sic] were plus knowing there could be a supply shortage. I realized later, just trust the software. that worked better''}--(PL117, male, 58)
\end{quote}

Reactors were also trying to ``keep safety stock'' ($n=5$), and ``stockpile'' ($n=3$), however, a dominant response in their comments was to ``balance inventory and backlog'' ($n=6$):

\begin{quote}
    \textit{``To minimize the losses incurred from the backlog while not keeping too much in the inventory...''}--(PL8, male, 29) \\
    \textit{``..I ordered less if I had a larger inventory, and more as my inventory began to dwindle, to try and balance out my backlog with how much I had to feed demand.''}--(PL29, male, 22)
\end{quote}

Finally, an interesting finding was the reasons provided by some reactors for deviating from the suggestions after the disruption, which implied a feeling of ``regret'' ($n=3$): \textit{``..I should have ordered a lot more straight away rather than wait to see how things would work.''}--(PL42, male, 43).

\subsubsection{Followers}
The last group of players comprises a small fraction of the players (14\%, $n=17$). Followers consistently follow the order suggestions throughout the game and stay in the control mode with the highest probability compared to the other player types (see the right transition matrix in Figure~\ref{fig:hmm_transitions_for_player_types}). Figure~\ref{fig:hmm_player_types} shows that only three players in only a few occurrences are in positive or negative adjustment modes. In addition, the results from the chi-squared test of independence (see Table~\ref{tab:contingencyTable}) suggests that the frequencies of most response modes among followers are significantly less frequent, including negative adjustment modes of N2 ($p=.034$) and N1 ($p=.019$), as well as positive adjustment modes of P1 ($p=.003$), P2 ($p=.011$), P3 ($p=.009$), P4 ($p=.021$), and P5 ($p=.01$). We first suspected that followers were not engaged with the game and merely clicked through the experiment. However, after analyzing their interactions with the game's interface, time spent making decisions, and total playtime, we could not find grounds implying that these players were not engaged. Therefore, we considered their play pattern to be the result of their behavior. We think followers are very compliant with the order suggestion as they believed this would be the best strategy: 

\begin{quote}
    \textit{``I was following the instructions''}--(PL5, male, NA) \\
    \textit{``I followed the system instructions and the software policy.''}--(PL45, male, 39)
\end{quote}

\subsection{Characterizing the Interaction}\label{interaction_char}
We attempted to characterize the interaction of players with the simulation by focusing on the behavioral response modes in each phase of the system described in section~\ref{system_char}. Our goal is to find how the system states, player types, and our implemented manipulation (i.e., Info vs. No-Info) affected the interaction of players with the system. Figure~\ref{fig:response_modes_in_phases} illustrates the frequencies of players for each response mode over the system phases, experimental conditions, and for each player type. We considered each system phase separately and calculated the Pearson residuals of the chi-square test to find response modes that are more/less frequent.

\subsubsection{Phase 1}
Phase 1 includes Week 21-27 of the gameplay where the system is \textit{stable}. In this phase, both reactors and followers are in control mode. Therefore, we focus on and hoarders as they contain various response modes in their sequences (see Hoarders/Phase 1 in Figure~\ref{fig:response_modes_in_phases}). Pearson residuals of the chi-square test suggest that response modes corresponding to higher deviation levels are more frequent among hoarders who did not receive any information on their supplier inventory. In particular, response modes N2 ($p=.003$), P1 ($p=.032$), P3 ($p=.022$), and P4 ($p=.021$) are more frequent in hoarders in the No-Info group and response modes N1 ($p=.042$), P1 ($p<.001$) and P2 ($p=.012$) are more frequent in the Info group. Although one can argue that these players were experimenting with the game in Phase 1, this result suggests that information sharing can affect the behavior of hoarders by reducing their uncertainty and causing them to show less deviation from the order suggestions.

\subsubsection{Phase 2}
Phase 2 starts on Week 28 by delivering the news about the manufacturing shutdown to players and ends on Week 32. The system is still in a \textit{stable} state in this phase, and followers continue to stay in the control mode. Reactors start to react to the news about the disruption in this phase (see Reactors/Phase 2 in Figure~\ref{fig:response_modes_in_phases}). However, this transition out of the control mode is not significant enough to place reactors more frequently in other response modes compared to hoarders. Hoarders in the Info group show P1 more frequently in their sequences ($p=.013$), and reactors show P2 in their sequences less frequently ($p=.038$). The implication is that, to a great extent, both hoarders and reactors behave similarly in the face of uncertainties and regardless of sharing information.

\subsubsection{Phase 3}
\begin{figure*}[ht]
    \centering
    \includegraphics[width=0.85\textwidth]{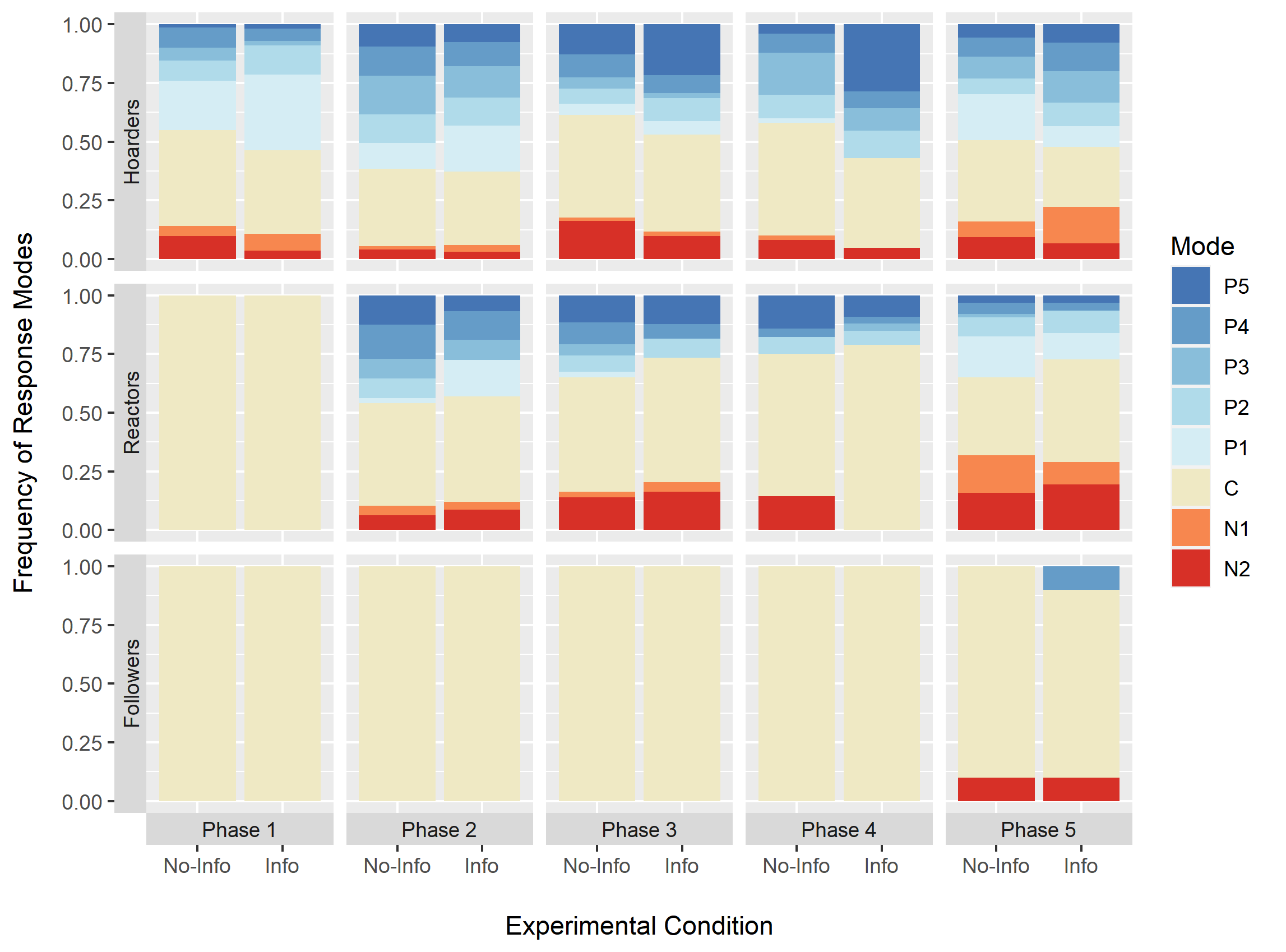}
    \captionsetup{width=\textwidth}
    \caption{Frequency of players who contained each response mode in their sequences across system phase and experimental conditions. An unexpected finding is that information sharing seems to affect the behavior of hoarders in Phase 4 where they show higher range of over-ordering (i.e., P5).}\label{fig:response_modes_in_phases}
    \Description[Frequency of response modes across five phases and for the three player types]{Stacked bar plots showing the frequency of response modes across the five phases of the system and the three player types. Bar plots represent the frequency of players who contained each response mode in their sequences across system phase and experimental conditions. An unexpected finding is that information sharing seems to affect the behavior of hoarders in Phase 4, where they show a higher range of over-ordering (i.e., P5).}
\end{figure*}

Phase 3 includes Weeks 33-36 and corresponds to the \textit{supply disruption} state. Most players experience the impact of manufacturing disruption during this phase by receiving fewer shipments from Manufacturer 1. In Phase 3, followers continue to stay in the control mode. At the same time, hoarders and reactors show positive and negative adjustment modes in their sequences (see Phase 3 in Figure~\ref{fig:response_modes_in_phases}). Pearson residuals of the chi-square test did not suggest any significant association between player types and information sharing in Phase 3. However, this is consistent with our previous finding, suggesting that both hoarders and reactors show the same level of uncertainty in facing disruptions regardless of sharing information. 

\subsubsection{Phase 4}
Phase 4 is a short period (Week 37-38); however, it marks the \textit{recovery from disruption} state where Manufacturer 1 starts sending shipments to Wholesaler 1 (i.e., the players' role). Players start to reduce the backlog they carried from the supply-disruption state. While followers are in control mode, hoarders and reactors show all response modes in their sequences. Pearson residuals of the chi-square test suggest P3 response mode to be significantly more frequent in the sequences of hoarders in the No-Info condition ($p=.007$), suggesting that hoarders tend to over-order even after \textit{supply disruption} phase. However, a more interesting finding is the P5 response mode, which is significantly more frequent among hoarders who received information on their supplier inventory ($p < 0.001$). Such behavior is unexpected for several reasons. First, these players could see their supplier inventory, knowing their supplier has recovered from the disruption. Second, the P5 mode corresponds to the highest level of deviation from the order suggestions (see Figure~\ref{fig:hmm_emissions}). While we expected hoarders to tend to over-order, we did not expect such behavior among the group who receives information sharing and especially during the \textit{recovery from disruption} state.

\subsubsection{Phase 5}
The final phase starts at Week 39, where for the majority of players, the system transits back to the \textit{stable} state. A few followers exhibit positive and negative adjustment modes; however, they mostly stay in the control mode. Reactors seem to show higher frequencies for negative adjustment modes (N1 and N2). Although the presence of such response modes in the sequences of reactors is not significantly more frequent than hoarders or followers, it suggests these players tend to under-order to reduce their excess inventory. Reactors in the Info condition do show significantly less frequency for P3 adjustment mode ($p=.038$). Interestingly, hoarders in the Info condition show N1 (i.e., slight negative adjustment) more frequently ($p=.022$), but also for positive adjustment modes of P3 ($p<0.001$) and P4 ($p=.01$). Hoarders continue to over-order, especially after recovering from uncertainties and when they have access to supplier inventory.

\section{Discussion}
Our results demonstrated the importance of characterizing the human-simulation interaction for advancing behavioral modeling and understanding of human decision-making. In what follows, we describe the main takeaways from our study by focusing on human-simulation interaction as a modeling approach when using game-based simulation environments. We further explain the most important aspects of our findings regarding the supply chain decisions.

\subsection{Characterizing Human-Simulation Interaction}
We attempted to characterize human-simulation interaction using the data collected from a gamette environment in an experimental study replicating a shortages scenario within a pharmaceutical supply chain. We also included a manipulation in our experiment where we shared supply chain information with some participants to test how information sharing affects human behavior. We focused on characterizing three main components involved in human-simulation interaction in our work. We first characterized the dynamic states of the system by using Principal Component Analysis (PCA) to reduce the dimensionality of the simulation and hierarchical cluster analysis to find similar system states. Second, we leveraged Hidden Markov Models (HMM) to characterize the behavioral responses of our human participants. Finally, we characterized the interaction of human players with the simulation by focusing on the behavioral responses across the system's states and considering our implemented manipulation. Below, we describe our main findings and the importance of characterizing each component.

\subsubsection{The Impact of the System}
In our characterization of the system, we found three states: \textit{stable}, \textit{supply disruption}, and \textit{recovery from disruption}. Each of these states represented specific situations in our supply chain simulation. While these states were mainly the result of our simulated shortage scenario, we showed how the behavior of individual players impacted the system state. For example, player PL77 never experienced the \textit{supply disruption} and \textit{recovery from disruption} states as the result of their decisions. Therefore, the system state affects human behavior, but players' behavior can also change the system's dynamic. Prior research provided evidence for similar reciprocal effects, especially in complex simulation environments where the individual decisions affect the system trajectories experienced by the player~\cite{sun2016modeling}. Gundry and Deterding~\cite{gundry2019validity} refer to such dynamics as \textit{variance}, which is one of the validity issues in the use of games for data collection in research. Removing variance by adding control contradicts the dynamicity of a simulation that aims to model real-world phenomena. Thus, it is important to have a system characterization as demonstrated in this paper. It can shed light on how different players experience the system, especially when using games as a medium for human-simulation interaction.

\subsubsection{The Impact of Human Behavior}
The main goal of many player modeling research is to characterize the cognitive, affective, and behavioral responses of human players~\cite{yannakakis2013player}. Using HMM, we found different behavioral response modes for how much players deviated from order suggestions in our experiment. In particular, we found \textit{positive}, \textit{control}, and \textit{negative} adjustment modes with different levels of deviations from the system's suggestion. Such behavioral responses can be partially attributed to the anchoring and adjustment heuristics~\cite{tversky1974judgment}. Prior research suggests humans make decisions by starting from an initial estimate (i.e., anchor) and then making adjustments for a final decision. The initial value can result from mental calculations or be suggested to the decision-makers (similar to the order suggestions in our experiment). In either case, the adjustments are expected to be insufficient and the final decision to be biased toward the anchor. 

Through sequence analysis of the behavioral response modes for each player, we found three types of players (hoarders, reactors, followers) based on how they adjusted the system's suggestions. In particular, hoarders showed to be more frequently in positive adjustment modes, while followers barely made any adjustments. As their name suggests, reactors reacted to the disruption news and made adjustments when faced with a shortage. This result is in line with previous work suggesting that some individuals have a higher tendency to hoard compared to others and that hoarding and panic buying are behavioral responses to scarcity~\cite{sterman2015m}. We can also study players' deviations from order suggestions from the perspective of trust in decision support recommendations. Prior research showed that trust plays an essential role in how humans adjust recommendations of decision support agents~\cite{wang2008attributions}. In addition, trust is a direct output of uncertainty~\cite{cho2015survey}. As a result, we expect deviations to be higher when players face uncertainties. In conclusion, we hypothesized that players deviate more from the system's suggestion during the shortage compared to a normal situation (H1) and found support for this hypothesis among hoarders and reactors. However, followers did not seem to lose trust in systems' suggestions when the supply chain was disrupted. While further research is warranted, especially into these followers, our results suggest that human behavior differs. 

\subsubsection{The Impact of the Manipulation}
We tested the effect of information sharing on players' behavioral responses by giving access to supplier inventory to some players. We used information sharing as our manipulation because it is considered an important strategy to mitigate disruptions and increase collaboration between stakeholders in supply chains~\cite{yaroson2019resilience, iyengar2016medicine}. While there are different types of information sharing such as upstream, downstream, or advanced information on drug shortages~\cite{pauwels2015insights}, we opted for upstream information sharing (i.e., supplier inventory) as it was the most straightforward to implement in terms of system manipulation. Investigating other types of information sharing is a point of interest in our future work. Prior work suggests information sharing helps reduce order fluctuations in the context of supply chains~\cite{yang2021behavioural}, and we hypothesized that information sharing reduces the amount of deviation from order suggestions (H2). While we found some evidence for this hypothesis for hoarders in Phase 1 of the simulation (i.e., the \textit{stable} state), our findings were entirely unexpected for the same group of players in Phase 4, which is when the players start to recover from the disruption. Hoarders who received information on their supplier inventory were more significantly in a high deviation response mode (P5) after the disruption period than those who did not receive this information. 

Such over-ordering behavior can be attributed to uncertainties about a future shortage. In addition, in our gamette, we framed the disruption scenario as a manufacturing shutdown due to the COVID-19 pandemic. We suspect that this behavior of hoarders, to some extent, could be related to their personal experiences of shortages in real life (i.e., toilet paper) and that they overestimated the demand in our game. However, it is important to note that this behavior was especially triggered with one type of players, the hoarders, and only when they received information---so certain people are more susceptible to over-ordering in certain contexts than others, again highlighting how human behavior differs. Prior work provides examples of similar situations in which stopping information sharing would be more beneficial by using system dynamics models~\cite{zhu2020managing}. Future research can extend current work by investigating such behaviors in an experimental setting and considering other forms of information sharing.

\subsection{Re-instituting Human-Simulation Interaction}\label{reinstitue_hsi}
Simulations are an abstracted version of reality~\cite{kunc2016behavioral}. However, most simulations are composed of a high dimensional space, especially when they are a model of complex systems such as supply chains. We showed how to reduce the dimensionality of a simulation by characterizing its different states. However, one might wonder why we need to characterize a simulation in the first place, especially one that we developed because we should be fully aware of its design. The answer is that we design simulations with predefined processes that can generate unexpected dynamics. Whether we model real-life processes (e.g., system dynamics), the flow of events or entities (e.g., discrete event simulation), or individuals' behaviors (e.g., agent-based simulation similar)~\cite{kunc2016behavioral}, we cannot easily predict the generated dynamics of any of these simulations due to their high dimensionality. The emerged states from our system characterization reflect the generated dynamics resulting from our simulation's predefined processes. In particular, we did not design our simulation with the \textit{stable}, \textit{supply disruption}, and \textit{recovery from disruption} states. These states are the outcomes of the flow of products within our system, summarizing what happened at specific points in time. Our results indicate that such characterizations are useful for making sense of the players' behavior and their interaction with the system. 

We used games as a medium for involving humans in the simulation. Games are complex systems and can generate a range of experiences that result in different behaviors. When coupled with a dynamic simulation, games can become even more complex, present cognitive overload, or introduce learning effects~\cite{gundry2019validity}. We have evidence for such complexities in our experiment where some players expressed ``frustration'' ($n=10$) because of the game dynamics. Others ``experimented'' ($n=6$) with the game to learn how it works. However, not all of our players shared the same experience. Others thought the game was ``easy'' to understand or control ($n=16$). The reason for such different experiences lies within how different players play a game. From the perspective of Triadic Game Design~\cite{harteveld2011triadic}, players play a game differently because of how they relate to the game (Reality), how they make sense of the game (Meaning), and how they play the game as players (Play). These are all reasons to avoid simplifying human behavior when modeling with data collected from game-based simulation environments. Although this has been the focus of player modeling for years, our goal is to shift the focus to how the interaction of humans with the environment forms human behavior. Therefore, when modeling player behavior, we need to consider the dynamic nature of the system as well as the intended manipulations. 

Manipulating the context to test hypotheses is one of the appeals of the use of game-based simulation environments~\cite{meijer2009organisation, mohaddesi_introducing_2020}. Gundry and Deterding~\cite{gundry2019validity} explained how any manipulation might result in unexpected interactions and emergent effects on player experience. Therefore, testing the effect of such manipulations in a dynamic environment may not be as straightforward. We showed how characterizing the interaction between players and the system allows us to investigate the effect of manipulations. Our proposed approach is one way to extract useful information from game-based simulation environments. It requires (1) characterizing the underlying simulation, (2) characterizing behavioral responses, and (3) characterizing the interaction between humans and the simulation while considering the manipulations. One can extend each step or use alternative techniques. For example, future research can further scrutinize the system characterization by decomposing the simulation into components, such as decomposing it into (1) the consequences of predefined processes in the simulation and (2) the dynamic changes of the system as the result of human behavior. For instance, in our characterization of the system states with PCA, a few players were characterized with a \textit{supply disruption} state before the actual disruption period (i.e., before week 28). While the \textit{supply disruption} states before and after week 28 share similar characteristics, they result from different dynamics. A decomposition allows us to better understand these observed outcomes by differentiating the states that may result from different dynamics. Alternative techniques to PCA such as Koopman Mode Analysis (KMA) and Dynamic Mode Decomposition (DMD) can also be considered, especially because these techniques are well-suited for characterizing temporal data~\cite{hogg2019koopman}.

Similarly, one can further scrutinize the characterization of players' behavior by analyzing observable responses (e.g., in-game actions) and non-observable responses (e.g., mental models, strategies, belief updates). In their review of player modeling literature, Snodgrass et al.~\cite{snodgrass2019like} point to various techniques for modeling player actions (i.e., observable), as well as affective states (i.e., non-observable). Deciding on which technique to use depends on the research question and the type of data available. We used HMM and sequence analysis, along with triangulating open response comments, as they suited our research question (i.e., hoarding in the context of drug shortages) and the sequential nature of decisions in our experiment. However, future research can leverage other techniques, such as inverse Bayesian inference for inferring behavioral tendencies~\cite{shergadwala2021can}. Alternatively, it is possible to complement behavioral data with biometric data, and apply Partially Observable Markov Decision Process (POMDP) and Bayesian belief updates as demonstrated by Chen et al.~\cite{chen2017cognitive}.

\subsection{Limitations}
We recruited online participants who did not necessarily have knowledge of supply chains. Therefore, some of the behaviors we observed could be associated with participants' lack of knowledge about how supply chains work. In our future studies, we plan to recruit supply chain experts and students with knowledge of supply chains to test the differences in their behavior. We also identified 14 players who were extreme outliers in their ordering decisions. While we decided to exclude these players from our analyses, their ordering strategy can be considered optimal under certain conditions. For example, some of these players ordered very large quantities and stopped ordering in subsequent weeks. Previous work also refers to situations in which the optimal ordering policy is to order one large quantity and then stop ordering for the subsequent periods~\cite{jain2017impact}. To scrutinize such behaviors, future research can focus on using methods that are robust to outliers~\cite{chatzis2008robust, rahmani2017coherence}. 

We also acknowledge that the difference between inventory and backlog cost in our simulation could have triggered the hoarding behavior in our participants. Players realized that facing backlog could cost them ten times more than holding extra inventory. We also only focused on deviations from order suggestions and did not include players' allocation decisions to reduce the complexity of our current analyses. Future research should analyze multi-dimensional decisions while characterizing human behavior in their interaction with the simulation. Finally, generalizing findings from game-based environments always requires certain scrutiny, especially given the inherent complexity of conducting research with games~\cite{gundry2019validity} that we just discussed. However, research has demonstrated that human behavior is by and large similar in games or virtual environments versus the real world~\cite{blascovich2011infinite}. Additionally, our work is based on~\cite{mohaddesi_introducing_2020} where the gamettes methodology was validated with regards to the established Beer Game.

\section{Conclusion}
In this paper, we proposed an approach for characterizing human-simulation interaction when using game-based simulation environments. We described how to characterize (1) the system state of the dynamic simulation environment, (2) behavioral responses from humans interacting with the simulation, and (3) characterizing the interaction by analyzing behavioral responses across system states. We provided empirical results from applying our approach to an experimental study on drug shortages in the context of supply chains. We found different player types (hoarders, reactor, follower) and showed how information sharing could exacerbate hoarding behavior after a period of shortage. Our work helps advance human-simulation interaction in characterizing different aspects of such environments and sheds light on the role of human behavior and decision-making. The latter is pertinent for addressing real-world problems because just as new drug shortages are bound to happen again, one and a half years from the start of the COVID-19 pandemic, people are stocking up on toilet paper again~\cite{terlep_americans_2021}.

\begin{acks}
This research was supported with funding from the National Science Foundation (NSF: 1638302). We further thank the StudyCrafter team and the Drug Shortage team at Northeastern University.
\end{acks}

\bibliographystyle{ACM-Reference-Format}
\bibliography{references}

\appendix
\section{Optimal Number of Clusters for Classifying Players}\label{appendix:optimal_clusters}
\begin{figure}[H]
    \centering
    \includegraphics[width=.5\columnwidth]{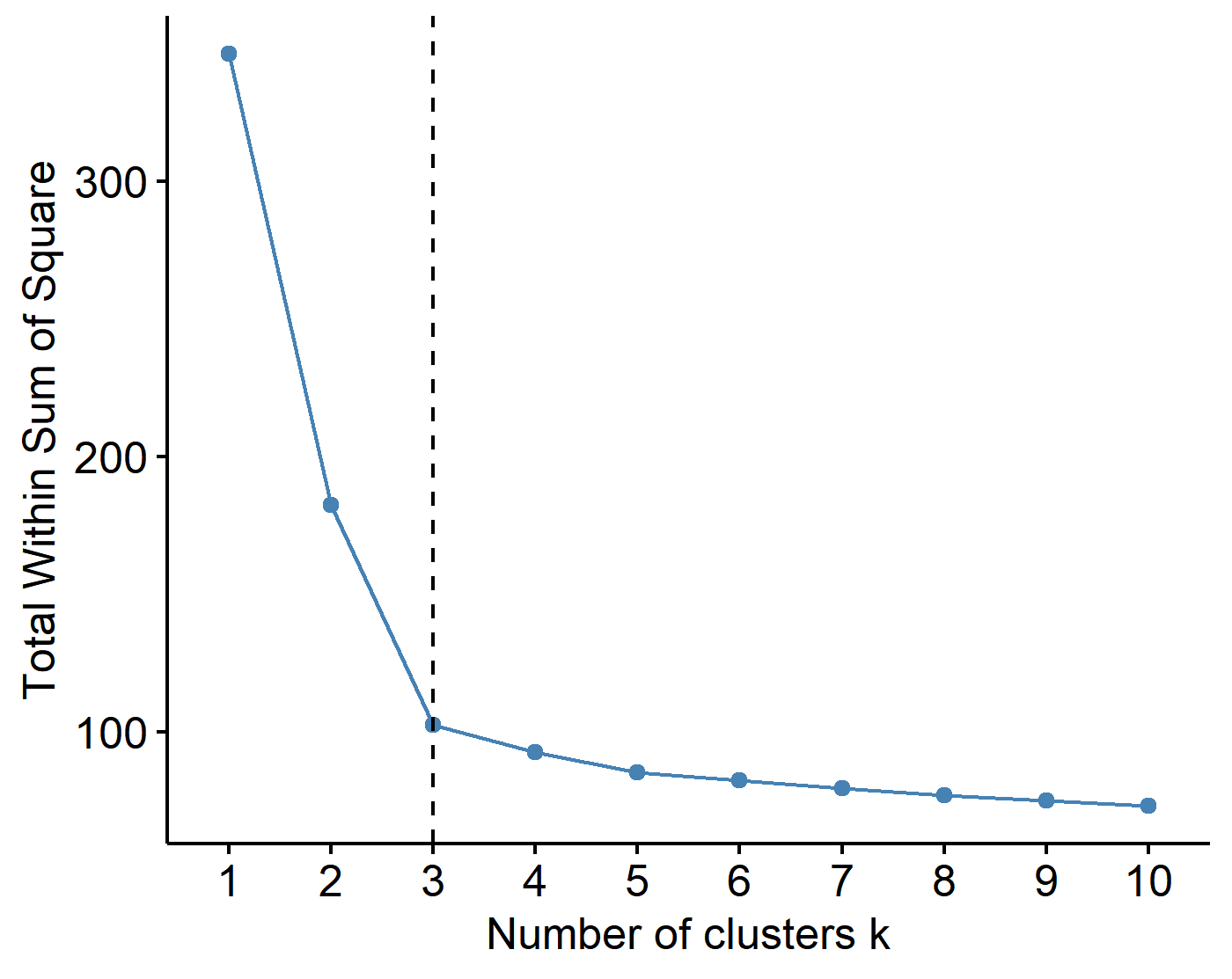}
    \captionsetup{width=\linewidth}
    \caption{Total within-cluster sum of square (WSS) calculated and plotted for different number of clusters. The plot suggests $k=3$ is optimal.}\label{fig:cluster_wss}
    \Description[Within-cluster sum of square values for different number of clusters]{Within-cluster sum of square values for different number of clusters where three clusters is shown to be optimal for classifying players into different types.}
\end{figure}

\begin{figure}[H]
    \centering
    \includegraphics[width=.5\columnwidth]{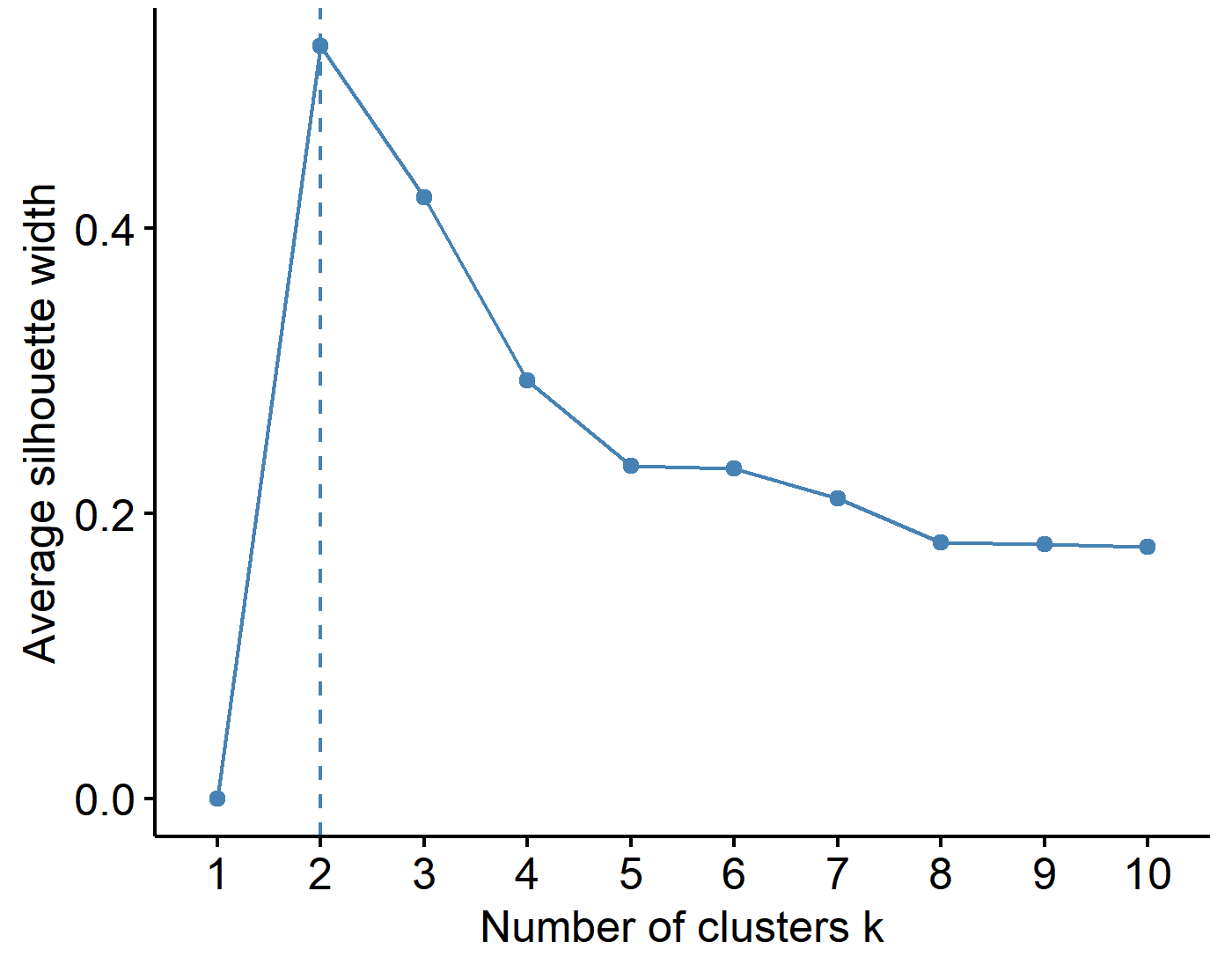}
    \captionsetup{width=\linewidth}
    \caption{Average silhouette width calculated and plotted for different number of clusters. The plot suggests $k=2$ is optimal.}\label{fig:cluster_silhouette}
    \Description[Average silhouette measure values for different number of cluster]{Average silhouette measure values for different number of cluster where two clusters is shown to maximize the average silhouette width indicating that two clusters is optimal for classifying players into different types.}
\end{figure}

\begin{figure}[H]
    \centering
    \includegraphics[width=.5\columnwidth]{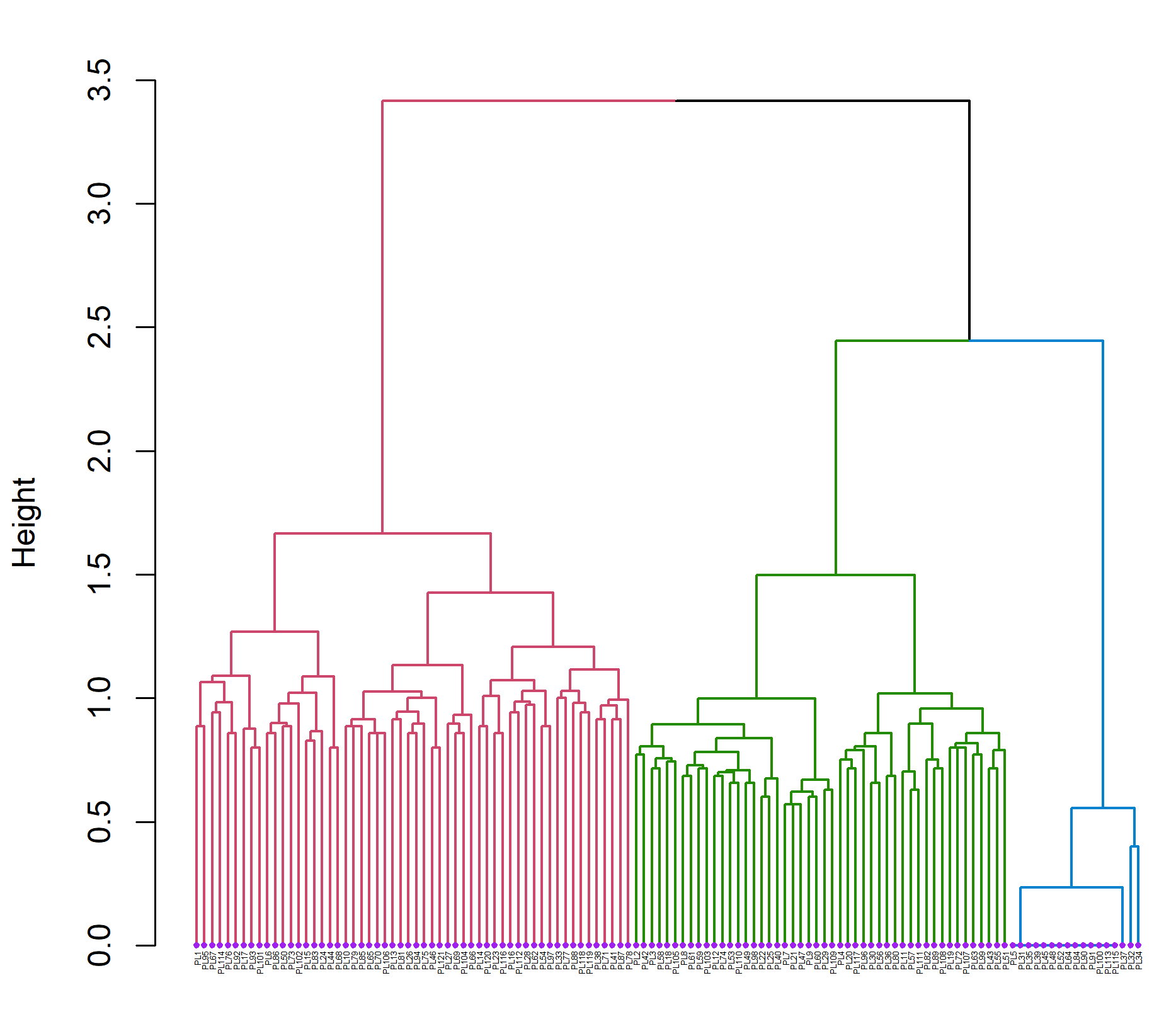}
    \captionsetup{width=\linewidth}
    \caption{Dendrogram representing the arrangement of players within each cluster and their relative distance according to hierarchical cluster analysis.}\label{fig:cluster_dendrogram}
    \Description[Arrangement of players within each cluster and their distance]{Arrangement of players within each cluster and their relative distance which shows three cluster can be used to classify players into different types.}
\end{figure}

To determine the optimal number of clusters for classifying players into different types, we first calculated the total within-cluster sum of square (WSS) and average silhouette width of observations for different values of $k$~\cite{kaufman2009finding}. While WSS suggested three clusters are optimal (see Figure~\ref{fig:cluster_wss}), silhouette measure suggested a two clusters solution (see Figure~\ref{fig:cluster_silhouette}). We also visually investigated the dendrogram of hierarchical cluster analysis (see Figure~\ref{fig:cluster_dendrogram}) and explored the player types with two and three clusters. Finally, We decided to choose $k=3$ as our optimal number of clusters as three clusters provided a better representation for different behaviors (see Figure~\ref{fig:hmm_player_types}).

\end{document}